\documentclass[sigconf]{acmart}
\usepackage{subcaption}
\usepackage{caption}
\usepackage[ruled,vlined,linesnumbered]{algorithm2e}
\usepackage{upquote}
\usepackage{listings}
\usepackage{comment}
\usepackage{todonotes}
\usepackage{fancyvrb}
\usepackage{graphicx}
\usepackage{multicol}
\usepackage{multirow}
\usepackage{cleveref}
\usepackage{lstautogobble}  
\usepackage{zi4}            
\usepackage{tabularx}
\usepackage{pgfplots,pgfplotstable}
\usepackage{soul}
\renewcommand\footnotetextcopyrightpermission[1]{}

\definecolor{greencomments}{rgb}{0, 0.5, 0}
\definecolor{redstrings}{rgb}{0.9, 0, 0}
\definecolor{graynumbers}{rgb}{0.5, 0.5, 0.5}
\definecolor{bluekeywords}{rgb}{0.13,0.13,1}

\definecolor{oricode}{RGB}{253, 231, 231} 
\definecolor{subcode}{RGB}{211, 235, 218} 
\definecolor{errotrigger}{RGB}{246, 237, 214} 

  {\list{}{\leftmargin=#1\rightmargin=#1}\item[]}%
  {\endlist}
\definecolor{darkgreen}{rgb}{0.0, 0.5, 0.0}
\newcommand{\rev}[1]{{#1}}
\usepgfplotslibrary{statistics}
\pgfplotsset{compat=1.8}

\begin{document}
\begin{sloppypar}
\title{PTE: Axiomatic Semantics based Compiler Testing}

\author{Guoliang Dong}
\email{gldong@smu.edu.sg}
\affiliation{%
  \institution{Singapore Management University}
  \city{Singapore}
  \country{Singapore}
}

\author{Jun Sun}
\email{junsun@smu.edu.sg}
\affiliation{%
  \institution{Singapore Management University}
  \city{Singapore}
  \country{Singapore}
}

\author{Richard Schumi}
\email{rschumi@smu.edu.sg}
\affiliation{%
  \institution{Singapore Management University}
  \city{Singapore}
  \country{Singapore}
}

\author{Bo Wang}
\email{wangbo_cs@bjtu.edu.cn}
\affiliation{%
  \institution{Beijing Jiaotong University}
  \city{Beijing}
  \country{China}
}

\author{Xinyu Wang}
\email{wangxinyu@zju.edu.cn}
\affiliation{%
  \institution{Shanghai Institute for Advanced Study of Zhejiang University}
  \city{Shanghai}
  \country{China}
}

\begin{abstract}
The correctness of a compiler affects the correctness of every program written in the language, and thus must be thoroughly evaluated. Existing automatic compiler testing methods however either rely on weak oracles (e.g., a program behaves the same if only dead code is modified), or require substantial initial effort (e.g., having a complete operational language semantics). 
\rev{While the former prevents a comprehensive correctness evaluation, the latter makes those methods irrelevant in practice.}
In this work, we propose an axiomatic semantics based approach for testing compilers, called PTE. \rev{The idea is to incrementally develop a set of ``axioms'' capturing anecdotes of the language semantics in the form of \emph{(\textbf{p}recondition, \textbf{t}ransformation, \textbf{e}xpectation)} triples, which allows us to test the compiler automatically.} Such axioms are written in the same language whose compiler is under test, and can be developed either based on the language specification, or by generalizing the bug reports. \rev{PTE has been applied to a newly developed compiler (i.e., Cangjie) and a mature compiler (i.e., Java), and successfully identified 42 implementation bugs and 9 potential language design issues.}   
\end{abstract}


\maketitle
\section{Introduction}
A bug in a compiler potentially renders all programs written in the language problematic. It is thus highly desirable that we have a systematic and scalable way of evaluating the correctness of compilers. The challenge of effective compiler testing~\cite{csmith,emi,sunLZS16,chen2020} is however enormous. A modern programming language often has many features which may evolve through time. That is, the compiler not only has to handle the complicated language semantics correctly, but also must keep up with the constant language updates. In addition, for efficiency reasons, a compiler is often teeming with various optimization options, which must be kept consistent with the evolving language semantics. All of this makes the task of evaluating the correctness of a compiler highly nontrivial. 

Based on the testing oracle, existing compiler testing approaches can be roughly categorized into three groups. The first group contains approaches that aim to maintain a comprehensive test suite~\cite{gcc,llvm,csmith}. \rev{This is certainly necessary, although it is hardly sufficient since covering every aspect of the compiler correctness} (e.g., the language semantics) would require a huge number of test cases. \rev{Creating and maintaining such a test suite is challenging, especially for evolving languages.}
 The second group contains approaches that aim to automatically test the compiler based on certain properties of the language semantics~\cite{ctyn2020}. One noticeable example is EMI~\cite{emi}, which is based on the property that altering `dead code' (i.e., part of the program that is not reachable given certain program input) should not alter program behaviors. While these approaches are shown to be effective in discovering compiler bugs~\cite{chen2018metamorphic}, they are often limited to simple algebraic properties. \rev{This means that they are far from comprehensively evaluating the correct implementation of the language semantics.} The third group contains those heroic efforts which aim to fully specify the (operational) semantics of a language and then evaluate the compiler's correctness accordingly~\cite{kjava,ksolidity,compcert}. While such approaches offer a way of systematically and comprehensively evaluating the correctness of the compiler, their adaptation in practice is hindered by the massive initial effort required to formalize the language semantics and the daunting task of maintaining the semantics subsequently. 

In this work, we aim to develop a compiler testing approach with the following objectives. First, it must be able to evaluate the compiler's correctness comprehensively. Second, it must not require a huge initial effort, and it must be relatively easy to maintain so that it can keep up with language updates. \rev{We thus propose an axiomatic semantics based approach called PTE (that stands for \textbf{p}recondition, \textbf{t}ransformation, and \textbf{e}xpectation) for testing compilers. The key idea behind PTE is to incrementally develop a set of axiomatic semantic rules which are used to systematically test the target compiler. Each axiomatic semantic rule is in the form of a triple \emph{(precondition, transformation, expectation)}, where \emph{precondition} specifies when the rule applies, \emph{transformation} describes a transformation of a program, and \emph{expectation} specifies the expected outcome of executing the transformed program.} For example, the PTE rule in Table~\ref{rule1} captures the Liskov Substitution Principle~\cite{liskov1987keynote} for object-orientation, which is often adopted as a language design principle. 
Each rule is implemented as a program in the language whose compiler is being tested. It is thus not necessary to learn a new language or notation. The actual implementation of a rule may have to resolve multiple details. \rev{For instance, for the high-level rule shown in Table}~\ref{rule1}, \rev{$o_2$ and $o_1$ must have the same values for the shared (static and instance) variables}. Furthermore, we must define the meaning of ``the program behavior is the same''.

\begin{table}[t]
\caption{PTE rule 1: `Liskov Substitution Principle' }
\label{rule1}
\begin{tabular}{c|l}
    P & There exists an object $o_1$ of type $Sup$ and \\
      & a class $Sub$ that is a subclass of $Sup$. \\
    \hline 
    T & Replace $o_1$ with an object $o_2$ of type $Sub$. \\
    \hline
    E & The program behavior remains the same. 
\end{tabular}
\end{table}
Given the implementation of a rule as well as a test suite for the compiler, PTE automatically identifies test cases which satisfy the \emph{precondition}; transforms the program according to the \emph{transformation}, and checks whether the \emph{expectation} is met by compiling and executing the transformed test cases with the target compiler.

While it is not hard to imagine that many such rules can be developed, in practice it is important to answer the question of how we can systematically develop, as well as maintain, a repository of such rules. We propose two ways of developing the rules. First, they can be built based on the language specification (written in natural languages). It is our observation that existing language specifications often contain `PTE rules' informally. For instance, here are some examples from the Java specification. 
\begin{itemize}
    \item \emph{``The enhanced for statement is equivalent to a basic for statement of the form: ...''}
    \item \emph{``It is a compile-time error if final appears more than once as a modifier for each variable declared in a resource specification.''}
\end{itemize}    
The former can be understood as a PTE rule whose \emph{transformation} is to de-sugar an enhanced for-statement to a basic one with the expectation of program equivalence. The latter can be turned into a PTE rule whose \emph{transformation} is to introduce an extra \emph{final} modifier with the expectation of a compile-time error. 

Second, we propose to incrementally build up the rule repository based on compiler bug fixes. We observed that it is often possible to develop PTE rules that generalize compiler bug fixes (see Section~\ref{sec:2}). 
Having such a way of developing and maintaining a rule repository is extremely important in practice, as it is more than a one-off effort with diminishing effect, but rather a way of systematically and incrementally approximating a language's semantics. 

To evaluate the feasibility and effectiveness of our approach, we have applied PTE to two programming languages, i.e., Cangjie and Java. These languages are chosen as representatives of a newly developed and a mature language. The experimental results show that PTE finds not only compiler bugs, but also potential issues regarding language design. \rev{In a short period of five months, we have defined and implemented dozens of PTE rules, and identified 40 unique bugs/issues in Cangjie and 11 bugs/issues in Java.} Most of them (2 issues in Cangjie and 3 issues in Java to be concluded) have been confirmed by the respective compiler team and 23 bugs in Cangjie have been fixed. 
While more PTE semantics rules are being developed continuously, we believe this result has successfully demonstrated the effectiveness of our approach. In fact, PTE has now been adopted by the Cangjie team who are working with us on building and maintaining the PTE rules repository. 

To sum up, we make the following technical contributions. First, we propose a novel, and importantly, practical compiler testing approach based on axiomatic semantics rules. \rev{Secondly, we define a set of PTE rules for Cangjie and Java and demonstrate their effectiveness by identifying previously unknown compiler bugs.}

\section{Illustrative Examples}
\label{sec:2}
\noindent{\textbf{\emph{Background.}}} \rev{Cangjie is a programming language that was newly developed by a leading global IT company, and it is currently undergoing rapid development. While the compiler team maintains a manually-written test suite (with about 20000 test cases), there is still a demand to comprehensively evaluate the correctness of the compiler. The idea of developing a formal language semantics (similarly as for the Solidity and the Java compiler}~\cite{ksolidity,kjava}) \rev{and a corresponding testing engine 
was proposed, but rejected due to the huge initial effort as well as the fact that it is almost impossible for the semantics to keep up with the rapid language development.} While multiple existing compiler testing techniques from the literature have been applied, such as EMI~\cite{emi} and SynFuzz~\cite{chen2023}, it is observed that such techniques often have a diminishing effect, i.e., they are able to find some bugs initially but soon become ineffective. More importantly, it is generally agreed upon by the testing team that many aspects of the language semantics are yet to be tested.

\rev{We propose PTE as a complementary way of testing the compiler. A set of rules in the form of \emph{(precondition, transformation, expectation)} triples are gradually developed, where \emph{precondition} is the condition under which the rule applies, \emph{transformation} specifies how a program is transformed, and \emph{expectation} is the expected outcome (e.g., the program finishes in the same program state or the same error is produced}).
These rules are gradually developed in two ways. One is based on the language specification (in both Chinese and English). The other is based on bugs that were raised and fixed, which are documented in the compiler project repository. That is, we examine each reported bug and see whether a rule can be developed to prevent such or similar issues in a generalized way. 

These rules are implemented in the form of a Cangjie program, utilizing existing facilities such as macros and a library for abstract syntax tree (AST) analysis (see Section~\ref{sec:3}). We show how PTE works using two example rules, one derived from the language specification and one from a bug report. 
\rev{We remark that all the bugs discussed below were discovered by our approach, i.e., they were not detected by existing approaches including manual testing and multiple automatic compiler testing methods.} \vspace{3pt}

\noindent{\textbf{\emph{PTE rules from language specification.}}} 
The Cangjie language specification~\cite{cjguide} includes a program (see below) with a variable initialization using a conditional expression. 
\begin{lstlisting}[numbers=none]
	let num = 8; let r = if (num > 0) { 1 } else { 0 }
\end{lstlisting}
\rev{This example gave us the idea that we can always replace a literal value with a conditional expression that is equivalent to the literal. We thus formulate a PTE rule shown in Table}~\ref{tab:ruleCE}, \rev{which states that, for an assignment, the value of the variable should remain unchanged if we rewrite the expression $v$ on the right side of `=' as 
a constant conditional expression whose result is always $v$.} 

\begin{table}[t]
  \caption{PTE rule 2: the conditional expression identity rule}
  \label{tab:ruleCE}
  \begin{tabular}{c|p{7cm}}
      P & The program has a statement which is either an assignment or a variable  declaration with an initialization. \\
      \hline 
      T & Replace the \emph{$v$} on the right of `=' with \emph{if(true)\{$v$\}else\{$v$\}}, where \emph{$v$} can be a variable, literal, or expression. \\
      \hline
      E & The program remains compilable and behaves the same. 
  \end{tabular}
\vspace{-5pt}
\end{table}
This PTE rule is implemented using existing Cangjie facilities (i.e., macros) and systematically applied to a suite of 5641 test cases for the Cangjie compiler. 
Out of which, 190 fail. After examining the failures manually, multiple bugs are identified. Figure~\ref{fig:r2} shows a simplified Cangjie program where the original code snippet is highlighted in red and the corresponding transformed code (after applying the rule) is marked in green. 
The transformed program is expected to be equivalent to the original one. However, compiling it 
triggers a crash. The error message says \emph{``Internal Compiler Error: Semantic error(s) in IR.''}, indicating that the compiler fails to generate a valid intermediate representation. This issue has been reported, confirmed and fixed in the Cangjie version 0.37.2. 

\begin{figure}[t]
    \begin{lstlisting}
      (*@\colorbox{oricode}{// var b ="" is Bool}@*)
      (*@\colorbox{subcode}{var b = if(true)\{"" is Bool\}else\{"" is Bool\}}@*)
      main(): Int64 {
        println(b)
        return 0
      }
    \end{lstlisting}
    \caption{The test case to which the rule in Table~\ref{tab:ruleCE} is applied}
    \label{fig:r2}
\end{figure}    

While it is not hard to find anecdotes from the language specification to develop PTE rules, in practice, the testing team is constantly worried about whether sufficiently many PTE rules have been developed, which is a relevant but challenging question. 
To answer it, we would have to check if the axiomatic semantics is equivalent to the operational language semantics, but this cannot be done without first developing the complete operational semantics.
A practical answer is that we can often develop new PTE rules based on bug reports, as we illustrate below. \vspace{3pt}

\begin{table}[t]
  \centering
  \small
  \caption{A Cangjie bug report~\cite{cjbugrp}}
  \label{tab:issue}
  \begin{tabular}{p{1.3cm}|p{5.5cm}}
    Title & Unexpected ``\{\}'' when using ``toTokens()''\\
    \hline
    Description & Unexpected ``\{\}'' when using ``toTokens()'' function to obtain tokens from an AST node of an interface which has an abstract property. \\
    \hline
    Steps of Reproduction & 
    \vspace{-3mm}
      \begin{lstlisting}[numbers=left,numbersep=5pt]
        main():Unit{
          let input = quote(interface A{ prop let a:Int64})
          let nodeA = parseInterfaceDecl(input)
          println(nodeA.toTokens())
      }
      \end{lstlisting}
    \\
    \hline
    Expected & \texttt{interface A \{ prop let a: Int64 \} } \ \\
    \hline
    Actual & \texttt{interface A \{ prop let a: Int64\textcolor{redstrings}{\{\}} \} } 
  \end{tabular}
\end{table}

\noindent{\textbf{\emph{PTE rules from bug reports.}}} Table~\ref{tab:issue} summarizes a bug reported from the Cangjie's forum. \rev{The bug affects multiple functions in the Cangjie core library which are designed to construct an AST from a Cangjie program and transform it into tokens}. The bug report shows a failed test case, where an interface \emph{A} is declared with an abstract property \emph{a}. Two built-in functions \emph{quote} and \emph{parseInterfaceDecl} are then used to construct the AST node, which is transformed into tokens using the \emph{toTokens()} function. The output is different from what is expected, and importantly, not a compilable Cangjie program. \rev{The cause of the bug is that a property defined in the above syntax is de-sugared during parsing to have an empty \emph{getter} and \emph{setter} declaration, represented as ``\{\}'' internally.} 

\begin{table}[t]
\caption{PTE rule 3: the toTokens-parse inverse rule}
\label{ruleInver}
\begin{tabular}{c|p{0.43\textwidth}}
    P & The program has a statement \emph{node.toTokens()} where \emph{node} is a variable or literal.  \\
    \hline 
    T & Replace the statement with \emph{parse(node.toTokens()).toTokens()} where \emph{parse} is a parsing function in Cangjie. \\
    \hline
    E & The program remains compilable and behaves the same. 
\end{tabular}
\end{table}

This bug is fixed in Cangjie version 0.37.2. 
\rev{Based on this bug, we formulate a generalized PTE rule (shown in Table}~\ref{ruleInver}\rev{), which intuitively says that parsing a Cangjie program (in the form of tokens) and then generating the tokens would give us back a compilable program which is equivalent to the original tokens.} We remark that this rule is not mentioned in the language specification or the documentation. Nonetheless, such a rule is expected to be satisfied in many scenarios. For instance, the \emph{toTokens()} function might be used for implementing code instrumentation (for various static or dynamic code analysis tasks), in which case, the un-instrumented part of the program is expected to be unchanged.

This PTE rule is applied to 5641 test cases for Cangjie. Out of which, 
186 fail. After examining the failures manually, multiple bugs are identified. 
Figure~\ref{fig:tkCode} shows one of the programs that produced a fault.
\rev{The transformation takes place at Lines 8 and 13, where the variable \emph{n} is replaced with \emph{parseExpr(n.toTokens())}.} Executing the transformed program results in a compile-time error, i.e., a violation of the expectation. The AST of an if-expression consists of two parts, i.e., one or more if-blocks and one optional else-block. 
\rev{Our PTE rule is applied to both Line 8 (the if-block) and Line 13 (the else-block). An error occurs at Line 13 for the else-block when the program is compiled. The error message, \emph{``error: [libast]: Parsing Error in ParseExpr''}, implies that 
the type of the else-block node (obtained by \emph{expr.getElseBranch()} at Line 10) is different from the type of the if-block node though they should be the same. This bug is fixed in version 0.37.2.} 
In fact, this one PTE rule allowed us to identify 8 bugs in total. \rev{Furthermore, it is agreed upon that this rule should be applied for future versions of the compiler since de-sugaring may be applied to newly designed language features.}

\begin{figure}[t]
    \centering
    \begin{lstlisting}
      main(){
        let input=quote(
          if ( x == "e" ) { return true } else { return false }
        )
        let expr=parseIfExpr(input)
        for (n in expr.getIfBody()){
          (*@\colorbox{oricode}{//n.toTokens().dump()}@*)
          (*@\colorbox{subcode}{parseExpr(n.toTokens()).toTokens().dump()}@*)        
        }
        match (expr.getElseBranch()) {
          case Some(n) =>
          (*@\colorbox{oricode}{//n.toTokens().dump()}@*)
          (*@\colorbox{subcode}{parseExpr(n.toTokens()).toTokens().dump()}@*)      
          case None => ()
        }
        return 0
      }
    \end{lstlisting}
    \caption{The test case to which the rule is applied} 
    \label{fig:tkCode}
\end{figure}

\section{Our Approach}
\label{sec:3}
In this section, we introduce how PTE is designed and realized. All the examples and discussions in this section are based on our implementation in Cangjie, although it should be clear that the approach naturally extends to other languages such as Java. 
 
\subsection{Overall Design}
\label{sec:3.1}
\begin{figure}[t]
\centering
    \includegraphics[width=0.48\textwidth]{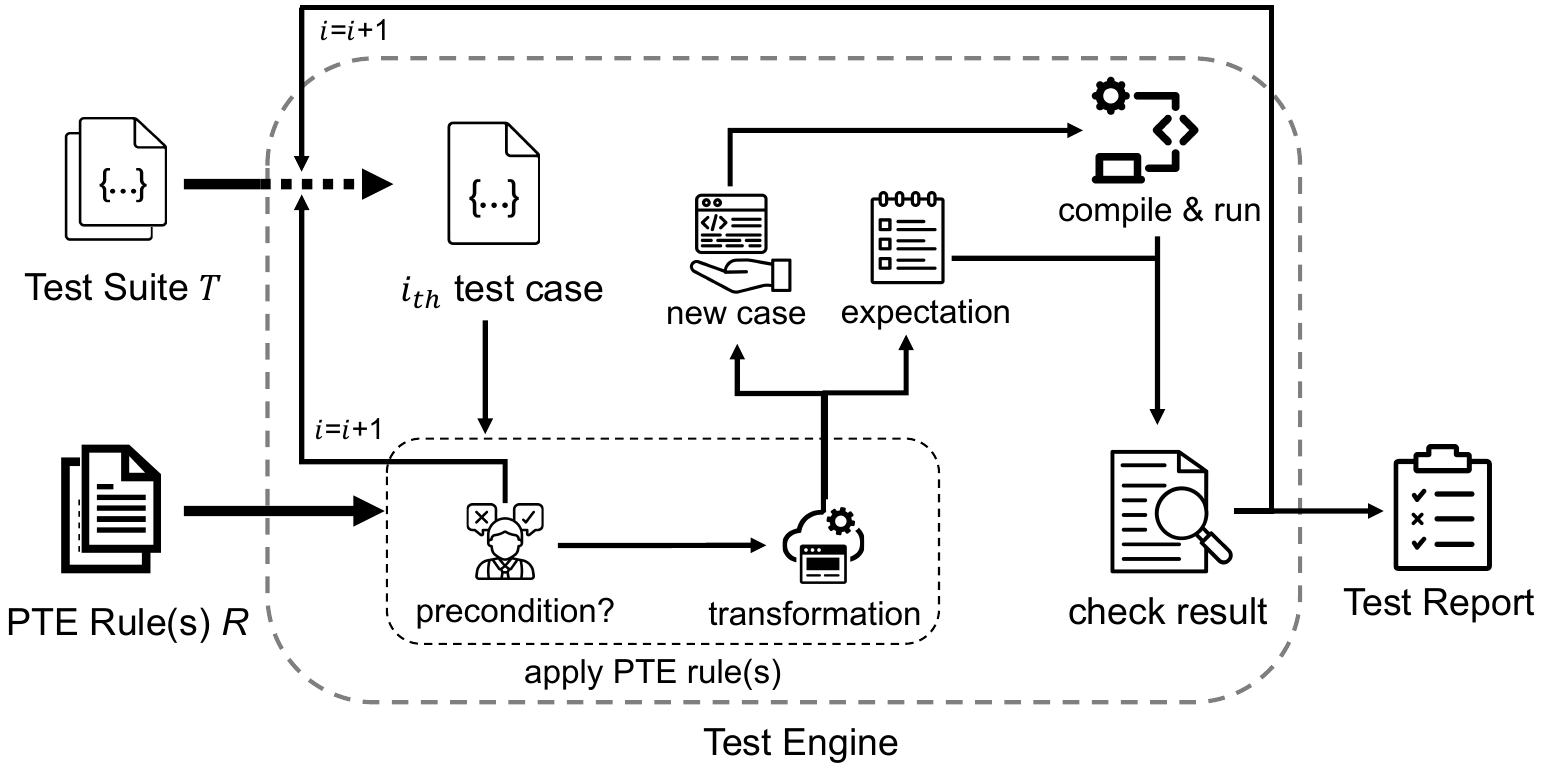}
    \caption{Overview of PTE}
    \label{fig:framwork}
\end{figure}

The overarching design principle of PTE is that it should be easy to apply. Having witnessed the difficulties of promoting (any kind of) formal notations and modeling in practice time and time again, we decide to design PTE such that a user does not need to learn any new language or notation to get started. Rather it should be possible to entirely rely on existing facilities provided by the language and the compiler. Furthermore, it should be possible to design and implement the rules incrementally and independently, i.e., a rule does not need to rely on other rules to work and it is not necessary to consider whether rules are overlapping or are complete. The reason is that we would like to accumulate a large number of PTE rules over time (so that PTE has a lasting effect) and it would be impossible to keep track of all the rules. \rev{Lastly, it should be easy to design and implement new rules or modify existing rules whenever a new language feature is introduced or the semantics of an existing feature is modified. 
An overview of PTE's simple design is shown in Figure}~\ref{fig:framwork}\rev{, which consists of three main components, i.e., an existing test suite $T$ for the compiler, a set of PTE rules $R$ and a test engine that takes $R$ and $T$.} 
In practice, such a test suite is often available, e.g., the OpenJDK test suite for Java contains over 10000 test cases~\cite{jdk} and the test suite for Cangjie contains about 20000 test cases.  
In the following, we describe the other two components in detail. 

\subsection{PTE Rules}
\label{sec:prec}
The core part of PTE is a repository of PTE rules, each of which implements the following interface.
\begin{lstlisting} [numbers=none]
    interface PTERule { 
        func precondition (program: Tokens) : Bool
        func transformation (program: Tokens) : Tokens 
        prop let expectations: ArrayList<Expectation> 
    }
\end{lstlisting}
A PTE rule has three parts, i.e., a precondition, a transformation and an expectation, which we describe in detail in the following. \\

\noindent \emph{\textbf{Precondition}.} The \emph{precondition} of a PTE rule determines if a rule is applicable to a program. It takes the form of a program written in the same language whose compiler is under test. The input of \emph{precondition} is a program (which has the type of \emph{Tokens} in Cangjie) and the output is either true or false. The precondition is usually implemented using existing meta-programming facilities. Note that different programming languages have varying degrees of support for meta-programming. Some programming languages provide built-in features specifically designed for meta-programming, whereas others require more advanced techniques to achieve similar results. For example, C++ uses templates to generate code at compile time, while Java relies on reflection to manipulate objects at runtime. Fortunately, Cangjie is designed explicitly with meta-programming in mind and offer macros that allow developers to manipulate code at compile time. Often, the \emph{precondition} of a PTE rule determines whether a PTE rule should be applied on the input program by checking whether the input program contains certain programming features. 
We refer readers to the Appendix for a detailed example on the implementation of one PTE rule. \\

\noindent \emph{\textbf{Transformation.}} \rev{Similarly, the \emph{transformation} of a PTE rule takes the form of a program written in the language whose compiler is to be tested, and it is also based on existing facilities for meta-programming.} The input of the program is a test program and the output is a transformed program. How the transformation is done is always specific to each PTE rule. 

We take the implementation of the transformation of the PTE rule shown in Table~\ref{tab:ruleCE} as an example. Given a program, we first convert it to an AST node, and then enumerate every one of its child AST nodes and replace the right side part of `=' in the assignment expression or variable declaration with the conditional expression. For a more complicated example, the  \emph{transformation} function of the PTE rule shown in Table~\ref{rule1} would depend on additional information, e.g., which classes are defined in the program and the inheritance relationship among these classes. \\ 
\begin{figure}
    \includegraphics[width=0.30\textwidth]{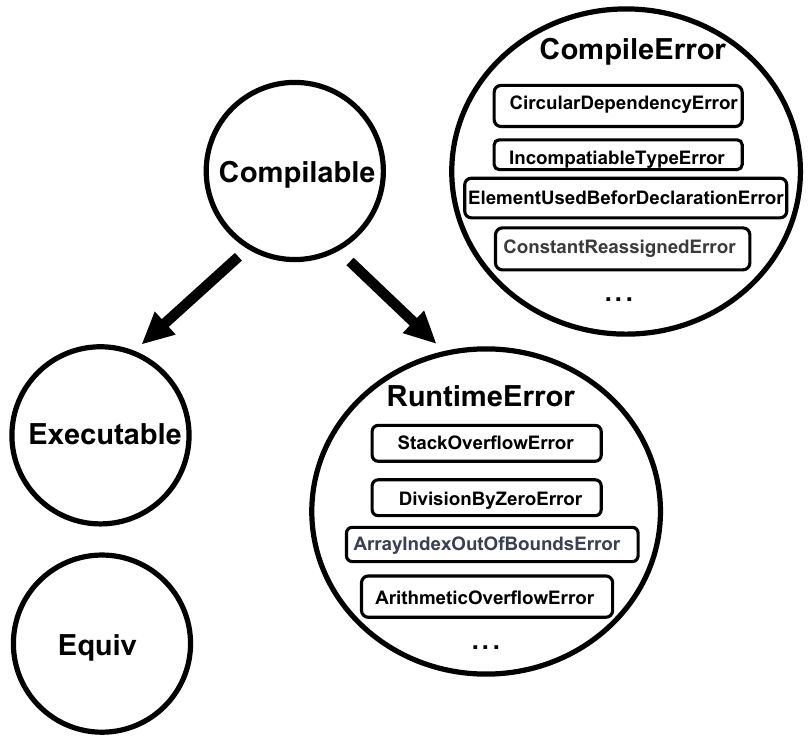}
    \caption{Different types of expectations}
    \label{fig:expeType}
\end{figure}

\noindent \emph{\textbf{Expectation.}} The \emph{expectation} of a PTE rule describes the expected behavior of the transformed program. Note that since a PTE rule is supposed to be applicable to any program satisfying the precondition, the expectation thus is mostly program-agnostic. In our work, we define different expectations for different stages of program compilation and execution, as shown in Figure~\ref{fig:expeType}. 

During compilation, we distinguish two kinds of expectations, i.e., $Compilable$ and $CompileError$. The former means that the program successfully compiles without any compile-time error, whereas the latter can be further detailed with specific compile-time errors. For instance, because Cangjie is designed to be strong-typed, assigning a String expression to a Boolean-typed variable is expected to cause a $IncompatiableTypeError$. Note that a compiler crash is never an expectation and thus would fail any expectation. During execution, we distinguish two expectations, i.e., $Executable$ and $RuntimeError$. The former means that the program successfully executes without any error, whereas the latter captures the expected runtime errors. For instance, we expect a $DivisionByZeroError$ whenever something is divided by zero. Exactly what kinds of compile-time and runtime errors are supported depends on the standardized error handling mechanism of the compiler. Lastly, we allow expectations that capture the relationship between the program before and after the transformation. For instance, $Equiv$ means that the program before and after the transformation has the same behavior, i.e., either both compilable and executable and having the same variable valuations (as well as user-output) at the end of the program execution, or resulting in the same compile-time error or runtime error. 


The expectations are implemented in Cangjie as an enum type. The default expectation is $Equiv$. All three PTE rules shown in the \Cref{rule1,tab:ruleCE,ruleInver} have the expectation of $Equiv$. Generally, each PTE rule is allowed to declare an array of expectations, which offers some flexibility in defining `imprecise' PTE rules. That is, ideally, a compiler tester should know precisely what the expected behavior of the compiler is when given a program. However, in practice, it might not be easy and it might simply be convenient to write PTE rules that have multiple expectations. For instance, a rule may state that for any program containing an Int64-typed variable $x$, introducing the code \emph{x-=1; x++} somewhere in the program after $x$ is initialized should either result in an equivalent program or an \emph{ArithmeticOverflowError} (since Cangjie is designed to conduct runtime overflow check for safety).

\subsection{Testing with PTE Rules}
\label{sec:testAlg}
\begin{algorithm}[t]
    \caption{PTE testing}
    \label{alg:pte1}
    \For{each test program $t_0 \in T$}{
        \For{each rule $r \in R$}{
            \If{$r.precondition(t_0)$}{
                let $t_1$ be $r.transformation(t_0)$\;
                $checkExpectation(r.expectations, t_0, t_1)$\;
            }        
        }
    }
\end{algorithm}
Given a test suite $T$ and a set of PTE rules $R$, a test engine decides how to apply the rules to the test suite. Algorithm~\ref{alg:pte1} shows a simple testing algorithm which applies each PTE rule to every test case in $T$ one-by-one. The function $checkExpectation$ takes the program before and after the transformation and checks whether the expectation is met according to the value of the property \emph{expectations}. For instance, if the expectation of the PTE rule is \emph{Equiv}, the test engine compiles and executes both the original program and the transformed program independently, records the information such as the output message and exit code for both programs, and compares them to determine if they are equivalent. If there are multiple expectations, we check each expectation one by one. As long as one of them is satisfied, it is considered that the expectation is satisfied. 

It is also possible and sometimes necessary to apply multiple PTE rules on the same test case. A practical example for this is that a compiler often applies multiple optimizations to generate more efficient executable code (e.g., the \emph{O3} option of GCC combines many optimizations). Given that each such optimization can be naturally encoded as a PTE rule (with the expectation \emph{Equiv}), our approach thus offers a way of checking whether applying multiple optimizations is safe or not. To test multiple PTE rules, we simply take a program and then apply the PTE rules one by one. That is, we first check whether the precondition of the first rule is satisfied, apply the transformation if it is, and then check whether the expectation is met. Afterwards, we apply the second rule based on the transformed program, and so on. We remark that this may also provide us with insights such as whether certain PTE rules are exclusive, i.e., applying a certain sequence of rules always renders certain rules inapplicable. 

\begin{figure}
    \centering
      \begin{lstlisting}[basicstyle=\footnotesize]
        open class Super { public var s1: UInt32 = 1 } 
        class Base <: Super { 
          public var b1: UInt32 = 2 
          (*@\colorbox{oricode}{// public var obj: Super = Super()}@*)
          (*@\colorbox{subcode}{public var obj: Super}@*)
          (*@\colorbox{subcode}{init()\{}@*) (*@\colorbox{subcode}{obj = Base()}@*) (*@\colorbox{subcode}{\}}@*)
        } 
      \end{lstlisting}
      \vspace{-2mm}
    \caption{An example of applying multiple rules}
    \label{fig:multi}
  \vspace{-6mm}
\end{figure}



Figure~\ref{fig:multi} shows an example which reveals a bug in the Cangjie compiler by applying two PTE rules. One is the rule shown in Table~\ref{rule1}, i.e., whose transformation replaces an object of class \emph{Super} with an object of its subclass \emph{Base}. The other is a rule which says that having a member variable initialized in the constructor (i.e., \emph{init()}) or in the containing class should be the same. As shown in Figure~\ref{fig:multi}, after applying both rules, the program is transformed such that Line 4 becomes Line 5 to 6. It turns out that the program before the transformation results in a \emph{StackOverflowError} whereas the transformed program results in a compile-time error (i.e., \emph{circular dependency}), which violates the \emph{Equiv} expectation.

\subsection{Developing and Maintaining PTE Rules}
\label{sec:3.6}
So far, we aim to convince readers that all it takes to apply PTE is to develop one PTE rule at a time, which is true except that developing and maintaining PTE rules could be non-trivial in practice. Such difficulties can however often be overcome by ``compromising'' the preciseness of the rule. In the following, we illustrate the complications in developing and maintaining PTE rules through examples. \\

\noindent \emph{\textbf{Developing PTE rules could be non-trivial.}} \rev{For instance, the rule shown in Table}~\ref{rule1} \rev{is one of the guiding principles of Cangjie, which says that objects of a superclass should be replaceable with objects of its subclasses.} Implementing such a rule for automatic compiler testing is however non-trivial as multiple critical details are missing from such a general description. For instance, how do we construct a corresponding instance of the subclass if additional state variables are needed? In our work, we develop an imperfect implementation of the rule (see Figure~\ref{fig:liskovTrans} in Appendix for details). Intuitively, if there is an expression of the form $\mathit{obj}.f()$ where $\mathit{obj}$ matches a class which has at least one qualified subclass, we then replace the base function name with the name of the subclass. A subclass is considered qualified if its constructor's signature matches the signature of the superclass. Note that a variety of additional information is required for the rule's implementation, i.e., we first traverse the program to obtain all the classes and their inheritance relationships, and then try to infer the type of the arguments in all constructor calls. 
For simplicity, we do not consider the scenario where additional state variables are required to instantiate the subclass. We assume that only the subclass whose constructor matches the superclass's constructor call can substitute the superclass. The expectation is set to be \emph{Equiv}.
\begin{figure}[t]
    \centering
    \begin{subfigure}[b]{0.15\textwidth}
        \begin{lstlisting}[basicstyle=\tiny]
        open class C1 { 
        public var x1:Int32=1 
        public open func f1(): Int32{
                return x1
            } 
        } 
        class C2 <: C1 { 
        public override func f1():Int32{ 
                return x1 + 1 
            } 
        } 
        main(): Int64 {
         (*@\colorbox{oricode}{// var v1 = C1().f1()}@*)
         (*@\colorbox{subcode}{var v1 = C2().f1()}@*)
          var v2 = C2().f1() 
          if(v1 == 1 && v2 == 2){ 
             return 0 
           } 
          return 1 
        } 
        \end{lstlisting}
        \caption{Polymorphism} 
        \label{fig:lspCase1}
    \end{subfigure}
    \begin{subfigure}[b]{0.15\textwidth}
        \begin{lstlisting}[basicstyle=\tiny]
            open class Mut{ 
              public func test(x:Int8):Int8{ 
                  return x 
                } 
            } 
            class subMut <: Mut { 
            public func test(x:Int16):Int16{ 
                    return x 
                } 
            }
            main(): Int64 { 
              (*@\colorbox{oricode}{//var rs=[subMut(),Mut()]}@*) 
              (*@\colorbox{subcode}{var rs=[subMut(),subMut()]}@*) 
              var r1 = ArrayList<Mut>(rs)
              return 0 
            } 
        \end{lstlisting}
        \caption{Array covariance} 
        \label{fig:lspCase2}
    \end{subfigure}
    \hspace{2pt}
    \begin{subfigure}[b]{0.15\textwidth}
    \begin{lstlisting}[basicstyle=\tiny]
      open class Super { 
        public var s1: UInt32 = 1 
      } 
      class Base <: Super { 
        public var b1: UInt32 = 2 
        (*@\colorbox{oricode}{//public var obj:Super=Super()}@*)
        (*@\colorbox{subcode}{public var obj:Super=Base()}@*)
      } 
      main() {     
        var mm: Base = Base() 
      } 
    \end{lstlisting}
        \caption{Circular dependency} 
        \label{fig:lspCase3}
    \end{subfigure}
  \caption{Violations of the Liskov Substitution Principle}
  \label{fig:lspCases}
  \end{figure}

\rev{The rule is then systematically and automatically applied to a suite of 5641 test programs for the Cangjie compiler, 26 of which fail.} After examining the failures manually, they are classified into three groups based on the underlying reasons for the failure. Figure~\ref{fig:lspCases} shows one example from each group. In particular, Figure~\ref{fig:lspCase1} shows a representative case (among 11 cases) where the failure is caused by polymorphism. The subclass \emph{C2} overrides the method \emph{f1} of superclass \emph{C1} and because \emph{f1} returns a different value in the two classes, a failure of the equivalence check occurs. Such failures are not considered bugs (but rather a design choice of many languages). Figure~\ref{fig:lspCase2} shows one of the two failures that are due to the fact that Cangjie does not support the array covariance. 
In particular, array covariance allows assigning an array of a subtype to a variable of an array type of its supertype, which is supported by many popular object-oriented languages, such as Java, C++ and C\#. 
After reporting this failure to the Cangjie team, we were informed that it is a design choice due to potential type safety concerns. 
The remaining 13 failures are due to a circular dependency, as exemplified in Figure~\ref{fig:lspCase3}. Originally, variable \emph{obj} is initialized to be an instance of the superclass. After applying the PTE rule and transforming the program, the variable is initialized to be an instance of the subclass, resulting in a circular dependency. The circular dependency in this case occurs because the initialization of class \emph{Base} depends on itself. This circular dependency escapes the compiler check somehow and incurs a \emph{StackOverflowError} during runtime. As discussed in Section~\ref{sec:testAlg}, combining this rule with another rule allows us to reveal a bug in the compiler. 

Given the above `false alarms', the test engineer must refine the PTE rule to avoid generating the same failures, e.g., by strengthening the precondition and adjusting how the transformation is done or weakening the expectation to cover such false alarms. For instance, to avoid the first group of failures, we need to check if an invoked called function (e.g, \emph{f1()}) is overwritten by the subclass using the precondition. Instead of strengthening the precondition using three conditions (i.e., one for each group of the failures), we can alternatively weaken the expectation, which is often easier in practice. Figure~\ref{fig:newExpeLiskov} shows the amended expectations. That is, we replace the \emph{Equiv} with a relaxed one, i.e., \emph{Executable}, to eliminate the false alarms caused by the polymorphism. Furthermore, we add two new expectations \emph{CircularDependencyError} and \emph{IncompatiableTypeError} to eliminate the false alarms caused by the circular dependency and the lack of support for array covariance. 

\rev{We remark that developing PTE rules is an iterative process in general. An initial implementation of a PTE rule would often be imperfect, e.g., it would result in multiple false alarms due to corner cases that are overlooked. These false alarms then serve as valuable references for refining the PTE rule and the implementation. Afterwards, we often re-run the PTE rule and analyze the reported failures, and refine the PTE rule further if necessary.} \\

\begin{figure}[t]
\begin{lstlisting}[basicstyle=\footnotesize]
            public override prop let expectations:Array<Expectation>{
              get(){
                (*@\colorbox{oricode}{//let equiv = Expectation(Equiv)}@*)
                (*@\colorbox{subcode}{let executable = Expectation(Executable)}@*)
                (*@\colorbox{subcode}{let circurlarE =Expectation(CircularDependencyError)}@*)
                (*@\colorbox{subcode}{let mismatchedE =Expectation(IncompatiableTypeError)}@*)
                return [executable,circurlarE,mismatchedE]
              }
            }
        \end{lstlisting}
    \caption{The refined expectations for the rule in Table~\ref{rule1}}
    \label{fig:newExpeLiskov}    
    \vspace{-3mm}
\end{figure}

\noindent \emph{\textbf{Maintaining PTE rules could be nontrivial.}} Ideally, PTE should be applied systematically such that over time a comprehensive set of rules are developed. To make it happen, we must be able to systematically maintain the rules so that they remain correct and applicable through language evolution. This is particularly relevant for newly developed languages such as Cangjie which constantly introduces new features or updates existing features, and not unimportant for mature languages such as Java. If a new language feature is introduced, our experience is that typically we can develop multiple new PTE rules to capture the axiomatic semantics of the new language feature. If however an existing feature is amended, identifying and updating those relevant rules that have become invalid may not be easy, as going through all the rules each time is impractical. In practice, however, this problem is manageable. Typically, whenever an existing language feature is amended, new test cases which reflect the correct concrete program behavior are introduced into the test suite, or some existing test cases are adjusted. 
\rev{Running those rules that have become invalid on these test cases will most likely result in failures, which makes it easy to identify and update such rules.}
For instance, PTE was initially implemented for Cangjie 0.34.3. Since then, Cangjie has evolved to 0.39.4 (i.e., the latest version at the time of writing), and many changes have been introduced. When we run PTE developed for Cangjie 0.34.3 on Cangjie 0.39.4, about 55 failures occur due to language changes. The causes of these errors are quickly identified (e.g., the `let' keyword is not needed when declaring a property, shadowing a function parameter with a local variable in the function is no longer allowed, and some built-in functions are no longer available) and the relevant rules are fixed within an hour.

\section{Implementation and Evaluation} \label{sec:4}
In this section, we present details of our experiments on applying PTE in practice. We have evaluated PTE for two compilers, one for Cangjie and one for Java (version 20). Based on the support for macros and the built-in AST library of Cangjie (and with the help of the Eclipse Java development tools (JDT) core library~\cite{jdtclib}), PTE is implemented with approximately 7500 lines of code for Cangjie and about 3000 lines for Java. Note that the Cangjie implementation additionally provides a range of APIs to facilitate developing PTE rules, and streamlines the testing. 

In the following, we report our experiences on applying PTE to Cangjie and Java, by focusing on the following research questions: \textbf{RQ1)} How effective is PTE in finding compiler bugs? \textbf{RQ2)} Which types of bugs can PTE find? \textbf{RQ3)} How much effort is needed to apply PTE in practice?
All the experiments are conducted on a laptop equipped with 11th Gen Intel(R) Core(TM) i7-1165G7@2.80GHz and 32 GB RAM, running Ubuntu 22.04.2 LTS (64 bit).

\subsection{Applying PTE for Cangjie}
For Cangjie, a set of 40 PTE rules are defined and then applied to test the compiler. In the following, we present the experimental results and answer the above-mentioned research questions. \rev{For a baseline analysis, we compare PTE with three approaches that have been recently developed for Cangjie, i.e., CJSmith, SynFuzz}~\cite{chen2023} \rev{and MetaFuzz}~\cite{chen2023}\rev{, which are shown to be more effective than previous approaches such as AFL}~\cite{afl}. These three approaches are based on different techniques. \rev{In particular, CJSmith, which follows the idea of Csmith}~\cite{csmith}\rev{, is provided by the Cangjie team, and it automatically generates random test cases according to the Cangjie grammar.} SynFuzz is a test case synthesis technique which synthesizes test cases by inserting code snippets into a randomly selected seed program. MetaFuzz is inspired by the approach known as EMI (i.e., equivalence modulo inputs~\cite{emi}) and generates test cases by performing semantic-preserving transformations on existing test cases. 
 \rev{We remark that for CJSmith and SynFuzz, the test oracle is whether a generated test case causes a compiler crash, whereas the test oracle of MetaFuzz is whether a mutated test case produces the same output as its original counterpart. Note that except CJSmith (which generates test cases from scratch), SynFuzz, MetaFuzz and PTE require a suite of seed programs}. We first run all the manually crafted test cases developed by the Cangjie team, and then keep those which pass the tests as the seed programs. There are a total of 5641 seed programs. The toolkit of both SynFuzz and MetaFuzz requires the seed programs to be in a specific format, and it can randomly generate such seed programs. For a fair comparison, we use this toolkit to generate an equal number of seed programs, i.e., 5641, in the required format, and then take them as the seed programs for SynFuzz and MetaFuzz.  \\

\noindent \textbf{\emph{RQ1: How effective is PTE in finding bugs in Cangjie?}} To answer this question, we run the test engine using the approach depicted in Section~\ref{sec:testAlg} systematically with the 40 PTE rules and with a timeout of 72 hours (the maximum testing time used in work~\cite{chen2023}) on Cangjie 0.34.3 which is the latest version at the time. The same timeout and version of Cangjie are used for both PTE and the baseline approaches. 
\begin{table}[t]
  \centering
  \caption{Bug finding statistics}
  \vspace{-3mm}
  \label{tab:rq1t1}
  \begin{tabular}{llll}
    \textbf{Approach} & \textbf{\#Test Cases} & \textbf{\#Bugs} & \textbf{Time (Hours)} \\
    \hline
    PTE      &225640 & $34^*$ &  59.3 \\ 
    \rev{CJSmith} &905949 & 0  &  72\\
    SynFuzz  & 30525 & 0  & 72 \\
    MetaFuzz & 41120 & 3   & 72  \\
    \hline
  \end{tabular}
  \\\emph{\footnotesize{$*$ potential design issues are not included, which are shown in Table~\ref{tab:bugtypes}}}
  \vspace{-5mm}
\end{table}

The results are summarized in Table~\ref{tab:rq1t1}. Our approach identified 34 (confirmed) unique bugs, all of which were previously unknown (we also found 6 potential design issues, which will be discussed in RQ2). At the time of writing, 23 of them have been fixed by the Cangjie team. The remaining bugs will be fixed in a future version. \rev{In contrast, MetaFuzz only identified three bugs, and the remaining two baselines, i.e, CJSmith and SynFuzz fail to identify any bugs.} Although MetaFuzz identified three bugs, we were unable to reproduce them in the later version (Cangjie 0.35.6). It is possible that these bugs were fixed as a side-effect of the version iteration.

 There are multiple reasons why these approaches fail to effectively identify bugs. \rev{First, the biggest reason is perhaps that CJSmith, SynFuzz and MetaFuzz had been previously applied and those bugs had already been fixed.} While it is true that these approaches may have been successful in the beginning, it certainly shows their diminishing effect, i.e., these approaches have a one-time benefit. Furthermore, if we refer to the effectiveness of these approaches when they were applied for the first time (i.e., reported in~\cite{chen2023}), \rev{we observed that when applying  CJSmith, SynFuzz, and MetaFuzz to the Cangjie compiler of version 0.24.5, a total of 15, 53, and 39 crashes/inconsistencies were identified, respectively.} However, when these approaches were applied to the Cangjie compiler of version 0.26.1, the numbers of crashes/inconsistencies dropped significantly to 9, 35 and 24, respectively. Note that not every crash/inconsistency is a unique bug, i.e., a total of 11 confirmed and unique bugs were found as reported in~\cite{chen2023}. Second, we observe that many test cases generated by SynFuzz and MetaFuzz are invalid. A close investigation shows that it is because these two approaches are implemented for an early version of Cangjie, i.e., version 0.26.1. Since the Cangjie grammar undergoes substantial changes from version 0.26.1 to version 0.34.3, and more importantly, SynFuzz and MetaFuzz are highly coupled with Cangjie syntax, maintaining these approaches (i.e., so that they can be applied for newer versions) is highly nontrivial. In comparison, PTE is more decoupled from the Cangjie syntax (i.e., it relies on the PTE rules, the tests and the AST library) and is easier to maintain. As long as the AST library is updated, the changes required to update PTE are minimal. For example, from version 0.34.3 to the latest version 0.39.4, only a few minor modifications are required, as discussed in Section~\ref{sec:3.6}. 

\rev{Table}~\ref{tab:rq1t1} \rev{also shows the number of test cases generated by CJSmith, SynFuzz and MetaFuzz. We can observe that CJSmith generated the highest number of test cases, reaching 905949, but found no bugs.} On the other hand, SynFuzz and MetaFuzz were less efficient, i.e., only generating 30525 and 41129 test cases, respectively, within 72 hours. In contrast, PTE generated 225640 (i.e., $5641$ test cases $\times$ $40$ rules) test cases within about 59 hours. 

\rev{To have a fair comparison with CJSmith, SynFuzz and MetaFuzz, we conduct an additional experiment on applying PTE to Cangjie 0.24.5. 
Due to the significant syntax and semantics change between version 0.24.5 and 0.34.3, we are only able to apply 10 of the PTE rules on Cangjie 0.24.5. With only 10 PTE rules, we identified 63 bugs, 
including 10 compiler crashes (reporting `Segmentation fault' or `CodeGen Error') and 3 miscompilations. 
We further analyze whether the bugs that we found in Cangjie 0.24.5 and that in 0.34.3 overlap. The results show that nearly 53\% (18 out of 34) bugs are newly discovered in Cangjie 0.34.3. In contrast, the baseline approaches only identified three bugs in Cangjie 0.34.3.}  \rev{Our experience is that we are typically able to find new bugs once we introduce new PTE rules.} \rev{This result shows that PTE will have a lasting effect if we continue to introduce new PTE rules, e.g., based on bug reports.}\\

\begin{table}[t]
    \small
    \caption{Different types of bugs found in Cangjie and Java}
    \label{tab:bugtypes}
        \begin{tabular}{p{5cm}cc}
        \textbf{Type} & \textbf{\#Cangjie}& \textbf{\#Java} \\
        \hline
        Compiler crash  & 2      &0\\ \hline
        Miscompilation  & 1    &0\\ \hline
        Problematic error messages & 4 & 7\\
        \hline
        Inconsistent error detection & 1 & 1 \\
        \hline
        Bugs in core libraries & 26 & 0\\
        \hline        
        Potential design issues & 6 & 3\\  
        \end{tabular}
        \vspace{-2mm}
\end{table}

\noindent \textbf{\emph{RQ2: Which types of bugs can PTE find in Cangjie?}} To answer this question, we categorize our findings into different categories in Table~\ref{tab:bugtypes}. A compiler crash happens during compilation and produces an error message such as \emph{``Internal Compiler Error''}. 
A miscompilation occurs if the compiler generates incorrect machine code, which either produces an incorrect execution result or causes a program crash with an error which is not related to the program itself. Figure~\ref{fig:rq2c1} shows such an example where the resulting machine code crashes with a segmentation fault. The original code at Line 2 is transformed to the code at Line 3 by the PTE rule shown in Table~\ref{tab:ruleCE}. Note that this bug and the bug shown in Figure~\ref{fig:r2} are different as they have different causes and fixes, even though they were detected by the same PTE rule. 

\begin{figure}[pt]
    \centering
    \begin{subfigure}[b]{0.25\textwidth}
        \begin{lstlisting}[basicstyle=\tiny]
            public struct Test { 
             (*@\colorbox{oricode}{//static var x=(1,0.1,true,'a',"a")}@*)
             (*@\colorbox{subcode}{static var x=if(true){(1,0.1,true,'a',"a")}else{(1,0.1,true,'a',"a")}}@*)  
             public func getX() { Test.x } 
            }
            main(){
             var (a_,b_,c_,d_,e_) = Test().getX()
            println("after:${a_},${b_},${c_},${d_},${e_}")
             return 0
            } 
        \end{lstlisting}
        \caption{The test case}
         \label{fig:rq2c11}
    \end{subfigure}
    \begin{subfigure}[b]{0.2\textwidth}
        \begin{lstlisting}[basicstyle=\tiny,columns=fullflexible,numbers=none, framesep=5pt]
            F unhandled signal. sig: 11, siginfo: 0x5620cff428b0, context: 0x5620cff42780, mutator's mutatorPage: 0x7fa9a9f47000, si_addr: 0x8!
            F unhandled signal. sig: 6, siginfo: 0x5620cff41330, context: 0x5620cff41200, mutator's mutatorPage: 0x7fa9a9f47000, si_addr: 0x3ea000c8b53!
            Aborted
        \end{lstlisting}
        \caption{The error message}
        \label{fig:rq2c12}
    \end{subfigure}
    \caption{An example revealing the bug of miscompilation}
    \label{fig:rq2c1}
\end{figure}

Compiler error messages are intended to help developers identify and fix issues in their code by providing meaningful information about the whereabouts and nature of errors. A problematic error message is one that is misleading or excessive, which makes it difficult for developers to identify and resolve the issue in the program. Figure~\ref{fig:rq2iem} shows such an example. In this example, the array \emph{ipmask} initially holds elements of type \emph{UInt8} (Line 2) and is transformed to the array which holds elements of type \emph{UInt64} by the rule (Line 3). Error messages such as \emph{mismatched types} are expected. However, the actual error message points the finger wrongly. It accuses an invalid subscript of the array \emph{ipmask}, whereas the subscript itself is valid and not the cause of the problem. This issue has been reported to the Cangjie team, and has been confirmed and fixed. The next category `inconsistent error detection' refers to a situation when a compiler behaves differently on equivalent programs in terms of bug detection. One such example has been shown in Figure~\ref{fig:multi}. The original test case has a circular dependency problem since the initialization of the class \emph{Base} depends on itself. This circular dependency escapes the check of the compiler and incurs a \emph{StackOverflowError} during runtime. However, after moving the initialization into the constructor, the compiler successfully detects the issue and generates a compile-time error. 
\begin{figure}[t]
    \begin{lstlisting}[basicstyle=\tiny]
        main() { 
           (*@\colorbox{oricode}{// var ipmask=Array<Array<UInt8>>([Array<UInt8>([255, 0, 0, 0]), Array<UInt8>([255, 255, 0, 0])]) }@*)
           (*@\colorbox{subcode}{var ipmask = Array<Array<UInt64>>([Array<UInt64>([255, 0, 0, 0]), Array<UInt64>([255, 255, 0, 0])]) }@*)
          for( i in 0 .. ipmask.size ){   var ipmask = IPMask(ipmask[i])} 
        } 
    \end{lstlisting}
    {\footnotesize \emph{error: invalid subscript operator [] on type `Struct-Array<Struct-Array<UInt64>>'}}
    \caption{An example showing problematic error message}
    \label{fig:rq2iem}
    \vspace{-6mm}
\end{figure}

Many of the detected bugs are in the Cangjie core library. These bugs range from minor issues that cause unexpected behavior to critical bugs that may lead to crashes or vulnerabilities. The following code reveals one such bug in the library for type conversion.
$$\mbox{\emph{let count=Int64(Float64.tryParse(0.toString()).getOrThrow())}}$$
The test case containing the above statement can be successfully compiled but encounters a \emph{NoneValueException} during runtime. The cause is that the function \emph{Float64.tryParse()} returns \emph{None} when converting string 0 to a number of \emph{Float64} type. The PTE rule involved in discovering the bug transforms an \emph{Integer} to a \emph{String}, then converts the \emph{String} to a floating-point number, and finally converts the floating-point number back to an \emph{Integer}. The precondition of this rule is that there exists an integer literal value which is assigned to a variable of type \emph{Int64}, and the expectation  is that the resulting integer value remains unchanged.

Besides implementation bugs, PTE also allows us to discover a number of potential design issues regarding the core libraries or the language itself. These flaws can result in inconvenience and confusion for Cangjie programmers, even though they may not be regarded as bugs. 
One example is that the exponentiation operation in Cangjie does not support negative exponent, which is an issue identified by the following transformed test program.
$$\mbox{\emph{let p=Int64.parse(2.toString())**Int64.parse(18.toString())}}$$
While there are reasons why it is designed so, negative exponents are a fundamental mathematical concept and are useful in various applications. Furthermore, it is supported by many programming languages such as Java and C++. 
Another example, as shown in Figure~\ref{fig:multiItf}, is that Cangjie lacks the ability to trace where a default implementation of a function in an interface is from. That is, if a class implements two interfaces, both of which inherits the same default function implementation from some common ancestor (i.e., another interface), a compile-time error occurs. The response from the Cangjie is that it is a design choice for now, i.e., Cangjie does not maintain the information on where a default implementation is from and has no way to check whether two implementations are in fact from the same source. The downside of such a design choice is that the programmers are forced to supply a new implementation which is likely a redundant copy of the same implementation. We thus consider such a design choice questionable and should be amended in the future. \\


\begin{figure}[pt]
    \centering
    \begin{subfigure}[b]{0.23\textwidth}
        \begin{lstlisting}[basicstyle=\tiny]
            interface Number < T > {   
                (*@\colorbox{subcode}{func cj1236ksjdnf13()\{ println( "hello!" ) \}}@*)
                operator func +( right: T ): T 
            } 
            interface Integer < T > <: Number < T > {} 
            extend Int32 <: Integer<Int64> & Number<Int8> { 
              public operator func +( right: Int64 ): Int64 { 3 } 
              public operator func +( right: Int8 ): Int8 { 4 } 
              public operator func +( right: Int16 ): Int16 { 5 } 
            } 
        \end{lstlisting}
        \caption{The test case}
         \label{fig:rq2c11}
    \end{subfigure}
    \begin{subfigure}[b]{0.23\textwidth}
        \begin{lstlisting}[basicstyle=\tiny,columns=fullflexible,numbers=none, framesep=5pt]
            error: interface function 'cj1236ksjdnf13' must be implemented in 'extend Int32'
            ==> main.cj:6:1:
               | 
             6 | extend Int32 <: Integer < Int64 > & Number < Int8 > { 
               | ^ 
               |  
            1 error generated, 1 error printed
        \end{lstlisting}
        \caption{The error message}
        \label{fig:rq2c12}
    \end{subfigure}
    \vspace{-3mm}
    \caption{A potential design issue}
    \label{fig:multiItf}
    \vspace{-3mm}
\end{figure}
\vspace{-1mm}
\noindent \textbf{\emph{RQ3: How much effort is needed to apply PTE for Cangjie?}} The effort required to apply PTE consists of multiple parts, e.g., manual efforts for implementing PTE rules and analyzing the testing results, and the computational resources for executing PTE. 

\rev{The time of developing PTE rules depends on the complexity of the rules and the facilities available for meta-programming. Empirically, the time for developing a PTE rule varies from 30 minutes to several hours using our framework. Most of the manual effort comes from implementing and refining the \emph{transformation} function of the PTE rule. For Cangjie, the transformations could be developed with build-in macros, and most of the rules (38 out of 40) are implemented with fewer than 100 lines of code. Unfortunately, Java does not support such macros, and hence an external AST processing library is required as we explain in Section}~\ref{sec:java}.

\rev{In total, it took 5 months to develop PTE for Cangjie, which includes time spent on learning Cangjie, developing the framework, designing/implementing rules based on the Cangjie language specification and its bug repository, and analyzing the results.
Note that once the framework is developed, adding new rules can be done efficiently. The Cangjie testing team managed to develop 30 rules in 3 days after PTE was adopted. 
The time to analyze the testing results varies depending on the nature and complexity of the issue. In our experience, most of the bugs are quickly identified and confirmed, whereas library/language design issues typically take much more time (sometimes multiple meetings spanning over weeks) to conclude.}


Regarding the computational resources for executing PTE, PTE can be run with a standard laptop or desktop computer. In our experiments, the average time is about 1.48 hours for each PTE rule (given the 5641 test programs). The total testing time for all PTE rules is about 59.3 hours. 
\rev{In contrast to alternative approaches such as CJSmith and SynFuzz which require a manually-specified timeout, PTE finishes when the test engine finishes all the test programs.} Furthermore, we can systematically measure the coverage of the PTE rules, which provides some test adequacy measure. 

\subsection{Applying PTE for Java}\label{sec:java}
In the following, we discuss our (brief) experiment on testing the Java 20 compiler with PTE. \\

\noindent \textbf{\emph{RQ1: How effective is PTE in finding bugs in Java?}}
Since Java is a mature language, most of its core language features have been intensively tested with various testing techniques and also manually by the large user base. We thus focus on the newly introduced features (i.e., preview or incubator features) in the latest Java version and develop 20 PTE rules based on the documentation. One such example is pattern matching for a switch or record patterns~\cite{javasr}.
We take those relevant tests in the openJDK test suite~\cite{jdk} as seed programs. Note that although openJDK consists of thousands of tests, most of them are considered irrelevant since our PTE rules are concerned with the newly introduced features. In the end, the PTE rules are applied to a range between 5 and 484 seed programs, depending on the type of the PTE rule.
We run the 20 PTE rules with the relevant seed programs as illustrated in Algorithm~\ref{alg:pte1}. 

\rev{We implement the PTE rules for Java by using an external library for AST parsing and modifying, i.e., the Eclipse Java development tools (JDT) core library which supports the latest Java version.}
Most of the effort in implementing PTE for Java is on realizing the AST modification with the JDT library, which takes hours to days depending on the complexity of the PTE rule. 
The average time on testing all the test cases is about 32min for each PTE rule. 

\rev{In total, we found 11 issues and bugs that were previously unknown. We reported the bugs to the Java bug database 
and eight of them have been confirmed (and will be fixed).} \\

\noindent \textbf{\emph{RQ2: What types of bugs can PTE find in Java?}} The types of bugs that we found for Java are summarized in the last of column of Table~\ref{tab:bugtypes}.
\rev{Seven bugs (bug ID: 8311136, 8308642, 8311135, 8308638, 8313437, 8313543, and 8313622) are related to problematic or misleading error messages.} For example (bug ID: 8308642), for the new switch feature with pattern matching, there occurs an unhelpful error message when a break statement is forgotten (see Figure~\ref{fig:javabug1} in Appendix). The error states \emph{``illegal fall-through to a pattern''}, which is unhelpful considering that most developers are not familiar with the fall-through behavior of a switch statement. Moreover, they may not know that the fall-through behavior is not allowed in the new pattern matching feature of Java 20. A more helpful error message should rather state that there is a missing break statement. The PTE rule that helped to find this bug is the one that introduces new redundant switch cases. 
\rev{Even though such error message related bugs might seem trivial, we believe it is important to fix them since they can cause a huge waste of debugging effort.}

We also found a case (bug ID: 8308636) of supposedly equivalent programs resulting in different compilation results. This bug is shown in Figure~\ref{fig:dominationerror} which includes a switch statement with the new guarded pattern feature~\cite{jdkguard}. The switch statement contains normal static cases, like case 1, and guarded pattern cases, like the second case. For this example, the Java compiler produces an error with the message \emph{``case label is dominated by a preceding case label''} for the third case, even though it is not the case. This error occurs due to an arbitrary design choice that requires static cases before pattern cases. Even if we agree with this design choice, the error message is unhelpful (and wrong) as a developer would not know about the required order of cases. 
\begin{figure}
	\centering
	\begin{lstlisting}[basicstyle=\footnotesize]
Integer i = 5;
switch (i) {
	case 1 -> System.out.println("Case 1!");   
	case Integer i1 when i1 < 0 -> System.out.println("Case 2!");             
	case 42 -> System.out.println("Case 3!");         
	default -> System.out.println("Default Case");
}
	\end{lstlisting}
        \vspace{-3mm}
	\caption{An example of wrong Java error message}
	\label{fig:dominationerror}
    \vspace{-6mm}
\end{figure}

We also found potential design issues with Java (bug ID: 8311134, 8308640, and 8311132). For example, for record classes, which are simple lightweight classes for storing data, we discover that a final modifier can be added, although it is redundant. This is because records are final by default, i.e., a record class cannot be extended. \rev{Allowing a final modifier may confuse developers, who might think that records without this modifier are extensible}.  

\section{Related work} \label{sec:5}

Compiler testing is a research topic that has attracted considerable interests. Broadly speaking, all approaches can be understood based on three key questions, i.e., the test case generation problem~\cite{hanford1970automatic}, the oracle problem~\cite{barr2014oracle}, and the test adequacy problem~\cite{zhu1997software}. For test generation, one common approach is to manually build and maintain a test suite. Major compiler projects like GCC, LLVM, and OpenJDK have their own manually created test suites~\cite{jdk,gcc}. These suites are designed to validate the compilers' functionality, correctness, and performance. However, relying solely on manually built test suites is insufficient. Achieving comprehensive coverage would require an overwhelming number of test cases, posing significant challenges in terms of creating and maintaining such test suites.

Researchers have explored the automatic generation of compiler test cases. Two popular categories of approaches are grammar-based approaches~\cite{guo2013automatic,purdom} and mutation-based approaches~\cite{papadakis2019mutation,lcm}. Grammar-based approaches typically start with a fixed code fragment, acting as a template or placeholder, and then use grammar rules to generate the remaining parts of the program. Csmith~\cite{csmith} is a well-known grammar-based tool based on a subset of the C grammar. It generates programs that cover a significant portion of C language features while avoiding undefined and unspecified behaviors. Sirer et al.~\cite{sirer1999using} propose test generation for JVM using production grammars. Some grammar-based approaches generate test cases solely based on the language grammar. For example, Purdom~\cite{purdom} proposed a sentence generator that traverses the productions of a set of context-free grammar rules. Zelenov and Zelenova~\cite{zelenov2006automated} present a method for producing test cases using a BNF grammar and a coverage criteria based on the target compiler's syntax analyzer. 

Mutation-based approaches generate programs by modifying existing test cases. For instance, LangFuzz~\cite{langfuzzing} takes a seed program and randomly replaces some non-terminals in the program with code fragments of the same type from a large code fragments pool. Another approach named JavaTailor~\cite{javatailor} mutates existing test cases to increase JVM code coverage. JavaTailor extracts various code fragments from historical bug-revealing test programs and randomly inserts them into the seed programs. EMI~\cite{emi} focuses on the `dead code' of the seed program, mutating an input program by deleting, inserting, or modifying unexecuted code.

With a collection of automatically generated test cases, the test oracle in compiler testing varies depending on the test case generation approach used. One popular approach is differential testing~\cite{mckeeman1998differential,ofenbeck2016randir,le2015randomized}, which aims to identify potential bugs by comparing the results of compiling and executing the same test case with different compilers or different optimization levels within a single compiler. \rev{For example, in the case of Csmith, each generated test case computes and prints a checksum of the non-pointer global variables, allowing the detection of discrepancies among different compilers}. Note that differential testing typically requires at least one mature and well-established compiler for the same language as a reference. However, this approach may not be directly applicable to newly developed programming languages, like Cangjie, where a mature compiler may not yet exist. 

Another approach to address the test oracle problem is metamorphic testing~\cite{ctyn2020,zhou2004metamorphic,segura2016survey}. Metamorphic testing focuses on defining metamorphic relations that specify how changes in the input program should affect the output. EMI serves as a typical example of metamorphic testing. The mutated programs generated by EMI are expected to produce equivalent results to their original counterparts when executed with a given set of test inputs. Bugs can be detected by comparing the results of the mutated program and its original counterpart. \rev{Donaldson et al.}~\cite{GLFuzz,GLFuzz1} \rev{introduced a metamorphic testing approach for graphics shader compilers that includes semantics-preserving mutations (e.g., like variable type replacement, expression reordering, or control flow wrapping) that affects code executed during testing.}
\rev{There are also similar metamorphic testing approaches}~\cite{HirGen,DBLP:journals/pomacs/XiaoLYPW22} \rev{for deep learning compilers.} 



\rev{Our approach can certainly benefit from existing approaches on compiler test generation such as grammar-based ones. The primary aim of our approach is to solve the test oracle problem. In particular, our approach falls under the category of metamorphic testing. However, unlike existing metamorphic testing approaches that typically focus on a single metamorphic relation, our approach provides a practical way for users to define metamorphic relations in the form of PTE rules. The diversity of these metamorphic relations enables us to uncover a broader range of language semantics and potential bugs. Our approach can be argued to be more general, i.e., it is designed to work with all types of compilers (not just domain specific ones), it supports transformations that are not semantic-preserving, and it decouples the oracle (i.e., the PTE rules) from the testing algorithm. Lastly, all the existing approaches mentioned earlier do not specifically address the test adequacy problem. In contrast, our approach has the potential capability to address this problem.  Theoretically, if our PTE rules cover all the semantics of the target language, we can claim that our approach is testing sufficiently. In the future, we aim to explore ways of measuring and approximating the coverage of language semantics based on the test cases and PTE rules, and further develop ways of improving such coverage through test generation.} 

\section{Conclusion} \label{sec:6}
In this work, we propose a new approach for compiler testing called PTE. The idea is to incrementally develop a set of axiomatic semantic rules in the form of \emph{(precondition, transformation, expectation)}, which subsequently are used to systematically test the target compiler. We applied our approach to two compilers, i.e., the Cangjie compiler and the Java compiler. \rev{Within a few months, we have identified 42 confirmed, unique bugs as well as 9 potential design issues in Cangjie and Java.} 

\bibliographystyle{ACM-Reference-Format}
\bibliography{ref}


\begin{thebibliography}{40}


\ifx \showCODEN    \undefined \def \showCODEN     #1{\unskip}     \fi
\ifx \showDOI      \undefined \def \showDOI       #1{#1}\fi
\ifx \showISBNx    \undefined \def \showISBNx     #1{\unskip}     \fi
\ifx \showISBNxiii \undefined \def \showISBNxiii  #1{\unskip}     \fi
\ifx \showISSN     \undefined \def \showISSN      #1{\unskip}     \fi
\ifx \showLCCN     \undefined \def \showLCCN      #1{\unskip}     \fi
\ifx \shownote     \undefined \def \shownote      #1{#1}          \fi
\ifx \showarticletitle \undefined \def \showarticletitle #1{#1}   \fi
\ifx \showURL      \undefined \def \showURL       {\relax}        \fi
\providecommand\bibfield[2]{#2}
\providecommand\bibinfo[2]{#2}
\providecommand\natexlab[1]{#1}
\providecommand\showeprint[2][]{arXiv:#2}

\bibitem[\protect\citeauthoryear{Barr, Harman, McMinn, Shahbaz, and Yoo}{Barr
  et~al\mbox{.}}{2014}]%
        {barr2014oracle}
\bibfield{author}{\bibinfo{person}{Earl~T Barr}, \bibinfo{person}{Mark Harman},
  \bibinfo{person}{Phil McMinn}, \bibinfo{person}{Muzammil Shahbaz}, {and}
  \bibinfo{person}{Shin Yoo}.} \bibinfo{year}{2014}\natexlab{}.
\newblock \showarticletitle{The oracle problem in software testing: A survey}.
\newblock \bibinfo{journal}{\emph{IEEE transactions on software engineering}}
  \bibinfo{volume}{41}, \bibinfo{number}{5} (\bibinfo{year}{2014}),
  \bibinfo{pages}{507--525}.
\newblock


\bibitem[\protect\citeauthoryear{Bogdanas and Rosu}{Bogdanas and Rosu}{2015}]%
        {kjava}
\bibfield{author}{\bibinfo{person}{Denis Bogdanas} {and}
  \bibinfo{person}{Grigore Rosu}.} \bibinfo{year}{2015}\natexlab{}.
\newblock \showarticletitle{K-Java: {A} Complete Semantics of Java}. In
  \bibinfo{booktitle}{\emph{Proceedings of the 42nd Annual {ACM}
  {SIGPLAN-SIGACT} Symposium on Principles of Programming Languages, {POPL}
  2015, Mumbai, India, January 15-17, 2015}},
  \bibfield{editor}{\bibinfo{person}{Sriram~K. Rajamani} {and}
  \bibinfo{person}{David Walker}} (Eds.). \bibinfo{publisher}{{ACM}},
  \bibinfo{pages}{445--456}.
\newblock


\bibitem[\protect\citeauthoryear{Chen, Patra, Pradel, Xiong, Zhang, Hao, and
  Zhang}{Chen et~al\mbox{.}}{2021}]%
        {chen2020}
\bibfield{author}{\bibinfo{person}{Junjie Chen}, \bibinfo{person}{Jibesh
  Patra}, \bibinfo{person}{Michael Pradel}, \bibinfo{person}{Yingfei Xiong},
  \bibinfo{person}{Hongyu Zhang}, \bibinfo{person}{Dan Hao}, {and}
  \bibinfo{person}{Lu Zhang}.} \bibinfo{year}{2021}\natexlab{}.
\newblock \showarticletitle{A Survey of Compiler Testing}.
\newblock \bibinfo{journal}{\emph{{ACM} Comput. Surv.}} \bibinfo{volume}{53},
  \bibinfo{number}{1} (\bibinfo{year}{2021}), \bibinfo{pages}{4:1--4:36}.
\newblock
\urldef\tempurl%
\url{https://doi.org/10.1145/3363562}
\showDOI{\tempurl}


\bibitem[\protect\citeauthoryear{Chen, Cheung, and Yiu}{Chen
  et~al\mbox{.}}{2020}]%
        {ctyn2020}
\bibfield{author}{\bibinfo{person}{Tsong~Yueh Chen}, \bibinfo{person}{S.~C.
  Cheung}, {and} \bibinfo{person}{Siu{-}Ming Yiu}.}
  \bibinfo{year}{2020}\natexlab{}.
\newblock \showarticletitle{Metamorphic Testing: {A} New Approach for
  Generating Next Test Cases}.
\newblock \bibinfo{journal}{\emph{CoRR}}  \bibinfo{volume}{abs/2002.12543}
  (\bibinfo{year}{2020}).
\newblock


\bibitem[\protect\citeauthoryear{Chen, Kuo, Liu, Poon, Towey, Tse, and
  Zhou}{Chen et~al\mbox{.}}{2018}]%
        {chen2018metamorphic}
\bibfield{author}{\bibinfo{person}{Tsong~Yueh Chen}, \bibinfo{person}{Fei-Ching
  Kuo}, \bibinfo{person}{Huai Liu}, \bibinfo{person}{Pak-Lok Poon},
  \bibinfo{person}{Dave Towey}, \bibinfo{person}{TH Tse}, {and}
  \bibinfo{person}{Zhi~Quan Zhou}.} \bibinfo{year}{2018}\natexlab{}.
\newblock \showarticletitle{Metamorphic testing: A review of challenges and
  opportunities}.
\newblock \bibinfo{journal}{\emph{ACM Computing Surveys (CSUR)}}
  \bibinfo{volume}{51}, \bibinfo{number}{1} (\bibinfo{year}{2018}),
  \bibinfo{pages}{1--27}.
\newblock


\bibitem[\protect\citeauthoryear{Donaldson, Evrard, Lascu, and
  Thomson}{Donaldson et~al\mbox{.}}{2017}]%
        {GLFuzz}
\bibfield{author}{\bibinfo{person}{Alastair~F. Donaldson},
  \bibinfo{person}{Hugues Evrard}, \bibinfo{person}{Andrei Lascu}, {and}
  \bibinfo{person}{Paul Thomson}.} \bibinfo{year}{2017}\natexlab{}.
\newblock \showarticletitle{Automated testing of graphics shader compilers}.
\newblock \bibinfo{journal}{\emph{Proc. {ACM} Program. Lang.}}
  \bibinfo{volume}{1}, \bibinfo{number}{{OOPSLA}} (\bibinfo{year}{2017}),
  \bibinfo{pages}{93:1--93:29}.
\newblock
\urldef\tempurl%
\url{https://doi.org/10.1145/3133917}
\showDOI{\tempurl}


\bibitem[\protect\citeauthoryear{Donaldson, Evrard, and Thomson}{Donaldson
  et~al\mbox{.}}{2020}]%
        {GLFuzz1}
\bibfield{author}{\bibinfo{person}{Alastair~F. Donaldson},
  \bibinfo{person}{Hugues Evrard}, {and} \bibinfo{person}{Paul Thomson}.}
  \bibinfo{year}{2020}\natexlab{}.
\newblock \showarticletitle{Putting Randomized Compiler Testing into Production
  (Experience Report)}. In \bibinfo{booktitle}{\emph{34th European Conference
  on Object-Oriented Programming, {ECOOP} 2020, November 15-17, 2020, Berlin,
  Germany (Virtual Conference)}} \emph{(\bibinfo{series}{LIPIcs},
  Vol.~\bibinfo{volume}{166})}, \bibfield{editor}{\bibinfo{person}{Robert
  Hirschfeld} {and} \bibinfo{person}{Tobias Pape}} (Eds.).
  \bibinfo{publisher}{Schloss Dagstuhl - Leibniz-Zentrum f{\"{u}}r Informatik},
  \bibinfo{pages}{22:1--22:29}.
\newblock
\urldef\tempurl%
\url{https://doi.org/10.4230/LIPICS.ECOOP.2020.22}
\showDOI{\tempurl}


\bibitem[\protect\citeauthoryear{Eclipse}{Eclipse}{2023}]%
        {jdtclib}
\bibfield{author}{\bibinfo{person}{Eclipse}.} \bibinfo{year}{2023}\natexlab{}.
\newblock
  \bibinfo{howpublished}{\url{https://github.com/eclipse-jdt/eclipse.jdt.core}}.
\newblock
\newblock
\shownote{Accessed Jun 26, 2023.}


\bibitem[\protect\citeauthoryear{GCC}{GCC}{2023}]%
        {gcc}
\bibfield{author}{\bibinfo{person}{GCC}.} \bibinfo{year}{2023}\natexlab{}.
\newblock \bibinfo{title}{7 Testsuites}.
\newblock
  \bibinfo{howpublished}{\url{https://gcc.gnu.org/onlinedocs/gccint/Testsuites.html}}.
\newblock
\newblock
\shownote{Accessed Jun 26, 2023.}


\bibitem[\protect\citeauthoryear{Guo and Qiu}{Guo and Qiu}{2013}]%
        {guo2013automatic}
\bibfield{author}{\bibinfo{person}{Hai-Feng Guo} {and} \bibinfo{person}{Zongyan
  Qiu}.} \bibinfo{year}{2013}\natexlab{}.
\newblock \showarticletitle{Automatic grammar-based test generation}. In
  \bibinfo{booktitle}{\emph{IFIP International Conference on Testing Software
  and Systems}}. Springer, \bibinfo{pages}{17--32}.
\newblock


\bibitem[\protect\citeauthoryear{Hanford}{Hanford}{1970}]%
        {hanford1970automatic}
\bibfield{author}{\bibinfo{person}{Kenneth~V. Hanford}.}
  \bibinfo{year}{1970}\natexlab{}.
\newblock \showarticletitle{Automatic generation of test cases}.
\newblock \bibinfo{journal}{\emph{IBM Systems Journal}} \bibinfo{volume}{9},
  \bibinfo{number}{4} (\bibinfo{year}{1970}), \bibinfo{pages}{242--257}.
\newblock


\bibitem[\protect\citeauthoryear{Holler, Herzig, Zeller, et~al\mbox{.}}{Holler
  et~al\mbox{.}}{2012}]%
        {langfuzzing}
\bibfield{author}{\bibinfo{person}{Christian Holler}, \bibinfo{person}{Kim
  Herzig}, \bibinfo{person}{Andreas Zeller}, {et~al\mbox{.}}}
  \bibinfo{year}{2012}\natexlab{}.
\newblock \showarticletitle{Fuzzing with Code Fragments.}. In
  \bibinfo{booktitle}{\emph{USENIX Security Symposium}}.
  \bibinfo{pages}{445--458}.
\newblock


\bibitem[\protect\citeauthoryear{Jiao, Kan, Lin, San{\'{a}}n, Liu, and
  Sun}{Jiao et~al\mbox{.}}{2020}]%
        {ksolidity}
\bibfield{author}{\bibinfo{person}{Jiao Jiao}, \bibinfo{person}{Shuanglong
  Kan}, \bibinfo{person}{Shang{-}Wei Lin}, \bibinfo{person}{David San{\'{a}}n},
  \bibinfo{person}{Yang Liu}, {and} \bibinfo{person}{Jun Sun}.}
  \bibinfo{year}{2020}\natexlab{}.
\newblock \showarticletitle{Semantic Understanding of Smart Contracts:
  Executable Operational Semantics of Solidity}. In
  \bibinfo{booktitle}{\emph{2020 {IEEE} Symposium on Security and Privacy, {SP}
  2020, San Francisco, CA, USA, May 18-21, 2020}}. \bibinfo{publisher}{{IEEE}},
  \bibinfo{pages}{1695--1712}.
\newblock


\bibitem[\protect\citeauthoryear{Le, Afshari, and Su}{Le et~al\mbox{.}}{2014}]%
        {emi}
\bibfield{author}{\bibinfo{person}{Vu Le}, \bibinfo{person}{Mehrdad Afshari},
  {and} \bibinfo{person}{Zhendong Su}.} \bibinfo{year}{2014}\natexlab{}.
\newblock \showarticletitle{Compiler validation via equivalence modulo inputs}.
  In \bibinfo{booktitle}{\emph{{ACM} {SIGPLAN} Conference on Programming
  Language Design and Implementation, {PLDI} '14, Edinburgh, United Kingdom -
  June 09 - 11, 2014}}, \bibfield{editor}{\bibinfo{person}{Michael F.~P.
  O'Boyle} {and} \bibinfo{person}{Keshav Pingali}} (Eds.).
  \bibinfo{publisher}{{ACM}}, \bibinfo{pages}{216--226}.
\newblock
\urldef\tempurl%
\url{https://doi.org/10.1145/2594291.2594334}
\showDOI{\tempurl}


\bibitem[\protect\citeauthoryear{Le, Sun, and Su}{Le et~al\mbox{.}}{2015}]%
        {le2015randomized}
\bibfield{author}{\bibinfo{person}{Vu Le}, \bibinfo{person}{Chengnian Sun},
  {and} \bibinfo{person}{Zhendong Su}.} \bibinfo{year}{2015}\natexlab{}.
\newblock \showarticletitle{Randomized stress-testing of link-time optimizers}.
  In \bibinfo{booktitle}{\emph{Proceedings of the 2015 international symposium
  on software testing and analysis}}. \bibinfo{pages}{327--337}.
\newblock


\bibitem[\protect\citeauthoryear{Leroy, Blazy, K{\"a}stner, Schommer, Pister,
  and Ferdinand}{Leroy et~al\mbox{.}}{2016}]%
        {compcert}
\bibfield{author}{\bibinfo{person}{Xavier Leroy}, \bibinfo{person}{Sandrine
  Blazy}, \bibinfo{person}{Daniel K{\"a}stner}, \bibinfo{person}{Bernhard
  Schommer}, \bibinfo{person}{Markus Pister}, {and} \bibinfo{person}{Christian
  Ferdinand}.} \bibinfo{year}{2016}\natexlab{}.
\newblock \showarticletitle{CompCert-a formally verified optimizing compiler}.
  In \bibinfo{booktitle}{\emph{ERTS 2016: Embedded Real Time Software and
  Systems, 8th European Congress}}.
\newblock


\bibitem[\protect\citeauthoryear{Liskov}{Liskov}{1987}]%
        {liskov1987keynote}
\bibfield{author}{\bibinfo{person}{Barbara Liskov}.}
  \bibinfo{year}{1987}\natexlab{}.
\newblock \showarticletitle{Keynote address-data abstraction and hierarchy}. In
  \bibinfo{booktitle}{\emph{Addendum to the proceedings on Object-oriented
  programming systems, languages and applications (Addendum)}}.
  \bibinfo{pages}{17--34}.
\newblock


\bibitem[\protect\citeauthoryear{LLVM}{LLVM}{2023}]%
        {llvm}
\bibfield{author}{\bibinfo{person}{LLVM}.} \bibinfo{year}{2023}\natexlab{}.
\newblock \bibinfo{title}{test-suite}.
\newblock
  \bibinfo{howpublished}{\url{https://github.com/llvm/llvm-test-suite.git}}.
\newblock
\newblock
\shownote{Accessed Jun 26, 2023.}


\bibitem[\protect\citeauthoryear{Ma, Shen, Tian, Chen, and Cheung}{Ma
  et~al\mbox{.}}{2023}]%
        {HirGen}
\bibfield{author}{\bibinfo{person}{Haoyang Ma}, \bibinfo{person}{Qingchao
  Shen}, \bibinfo{person}{Yongqiang Tian}, \bibinfo{person}{Junjie Chen}, {and}
  \bibinfo{person}{Shing{-}Chi Cheung}.} \bibinfo{year}{2023}\natexlab{}.
\newblock \showarticletitle{Fuzzing Deep Learning Compilers with HirGen}. In
  \bibinfo{booktitle}{\emph{Proceedings of the 32nd {ACM} {SIGSOFT}
  International Symposium on Software Testing and Analysis, {ISSTA} 2023,
  Seattle, WA, USA, July 17-21, 2023}},
  \bibfield{editor}{\bibinfo{person}{Ren{\'{e}} Just} {and}
  \bibinfo{person}{Gordon Fraser}} (Eds.). \bibinfo{publisher}{{ACM}},
  \bibinfo{pages}{248--260}.
\newblock
\urldef\tempurl%
\url{https://doi.org/10.1145/3597926.3598053}
\showDOI{\tempurl}


\bibitem[\protect\citeauthoryear{McKeeman}{McKeeman}{1998}]%
        {mckeeman1998differential}
\bibfield{author}{\bibinfo{person}{William~M McKeeman}.}
  \bibinfo{year}{1998}\natexlab{}.
\newblock \showarticletitle{Differential testing for software}.
\newblock \bibinfo{journal}{\emph{Digital Technical Journal}}
  \bibinfo{volume}{10}, \bibinfo{number}{1} (\bibinfo{year}{1998}),
  \bibinfo{pages}{100--107}.
\newblock


\bibitem[\protect\citeauthoryear{Ofenbeck, Rompf, and P{\"u}schel}{Ofenbeck
  et~al\mbox{.}}{2016}]%
        {ofenbeck2016randir}
\bibfield{author}{\bibinfo{person}{Georg Ofenbeck}, \bibinfo{person}{Tiark
  Rompf}, {and} \bibinfo{person}{Markus P{\"u}schel}.}
  \bibinfo{year}{2016}\natexlab{}.
\newblock \showarticletitle{RandIR: differential testing for embedded
  compilers}. In \bibinfo{booktitle}{\emph{Proceedings of the 2016 7th ACM
  SIGPLAN Symposium on Scala}}. \bibinfo{pages}{21--30}.
\newblock


\bibitem[\protect\citeauthoryear{OpenJDK}{OpenJDK}{2023a}]%
        {jdk}
\bibfield{author}{\bibinfo{person}{OpenJDK}.} \bibinfo{year}{2023}\natexlab{a}.
\newblock \bibinfo{title}{test-suite}.
\newblock
  \bibinfo{howpublished}{\url{https://openjdk.org/projects/code-tools/jtreg/intro.html}}.
\newblock
\newblock
\shownote{Accessed Jun 26, 2023.}


\bibitem[\protect\citeauthoryear{OpenJDK}{OpenJDK}{2023b}]%
        {jdkguard}
\bibfield{author}{\bibinfo{person}{OpenJDK}.} \bibinfo{year}{2023}\natexlab{b}.
\newblock \bibinfo{title}{test-suite}.
\newblock \bibinfo{howpublished}{\url{https://openjdk.org/jeps/406}}.
\newblock
\newblock
\shownote{Accessed Jun 26, 2023.}


\bibitem[\protect\citeauthoryear{Oracle}{Oracle}{2023}]%
        {javasr}
\bibfield{author}{\bibinfo{person}{Oracle}.} \bibinfo{year}{2023}\natexlab{}.
\newblock
  \bibinfo{howpublished}{\url{https://www.oracle.com/java/technologies/javase/20-relnote-issues.html}}.
\newblock
\newblock
\shownote{Accessed Jun 26, 2023.}


\bibitem[\protect\citeauthoryear{Papadakis, Kintis, Zhang, Jia, Le~Traon, and
  Harman}{Papadakis et~al\mbox{.}}{2019}]%
        {papadakis2019mutation}
\bibfield{author}{\bibinfo{person}{Mike Papadakis}, \bibinfo{person}{Marinos
  Kintis}, \bibinfo{person}{Jie Zhang}, \bibinfo{person}{Yue Jia},
  \bibinfo{person}{Yves Le~Traon}, {and} \bibinfo{person}{Mark Harman}.}
  \bibinfo{year}{2019}\natexlab{}.
\newblock \showarticletitle{Mutation testing advances: an analysis and survey}.
\newblock In \bibinfo{booktitle}{\emph{Advances in Computers}}.
  Vol.~\bibinfo{volume}{112}. \bibinfo{publisher}{Elsevier},
  \bibinfo{pages}{275--378}.
\newblock


\bibitem[\protect\citeauthoryear{PLLab}{PLLab}{2023a}]%
        {cjguide}
\bibfield{author}{\bibinfo{person}{PLLab}.} \bibinfo{year}{2023}\natexlab{a}.
\newblock
  \bibinfo{howpublished}{\url{https://gitee.com/HW-PLLab/cangjie/tree/master/docs/llvm}}.
\newblock
\newblock
\shownote{Accessed Jun 26, 2023.}


\bibitem[\protect\citeauthoryear{PLLab}{PLLab}{2023b}]%
        {cjbugrp}
\bibfield{author}{\bibinfo{person}{PLLab}.} \bibinfo{year}{2023}\natexlab{b}.
\newblock
  \bibinfo{howpublished}{\url{https://e.gitee.com/HW-PLLab-pro/issues/list?issue=I6F8S6}}.
\newblock
\newblock
\shownote{Accessed Jun 26, 2023.}


\bibitem[\protect\citeauthoryear{Purdom}{Purdom}{1972}]%
        {purdom}
\bibfield{author}{\bibinfo{person}{Paul Purdom}.}
  \bibinfo{year}{1972}\natexlab{}.
\newblock \showarticletitle{A sentence generator for testing parsers}.
\newblock \bibinfo{journal}{\emph{BIT Numerical Mathematics}}
  \bibinfo{volume}{12} (\bibinfo{year}{1972}), \bibinfo{pages}{366--375}.
\newblock


\bibitem[\protect\citeauthoryear{Segura, Fraser, Sanchez, and
  Ruiz-Cort{\'e}s}{Segura et~al\mbox{.}}{2016}]%
        {segura2016survey}
\bibfield{author}{\bibinfo{person}{Sergio Segura}, \bibinfo{person}{Gordon
  Fraser}, \bibinfo{person}{Ana~B Sanchez}, {and} \bibinfo{person}{Antonio
  Ruiz-Cort{\'e}s}.} \bibinfo{year}{2016}\natexlab{}.
\newblock \showarticletitle{A survey on metamorphic testing}.
\newblock \bibinfo{journal}{\emph{IEEE Transactions on software engineering}}
  \bibinfo{volume}{42}, \bibinfo{number}{9} (\bibinfo{year}{2016}),
  \bibinfo{pages}{805--824}.
\newblock


\bibitem[\protect\citeauthoryear{Sirer and Bershad}{Sirer and Bershad}{1999}]%
        {sirer1999using}
\bibfield{author}{\bibinfo{person}{Emin~G{\"u}n Sirer} {and}
  \bibinfo{person}{Brian~N Bershad}.} \bibinfo{year}{1999}\natexlab{}.
\newblock \showarticletitle{Using production grammars in software testing}.
\newblock \bibinfo{journal}{\emph{ACM SIGPLAN Notices}} \bibinfo{volume}{35},
  \bibinfo{number}{1} (\bibinfo{year}{1999}), \bibinfo{pages}{1--13}.
\newblock


\bibitem[\protect\citeauthoryear{Sun, Le, and Su}{Sun et~al\mbox{.}}{2016a}]%
        {lcm}
\bibfield{author}{\bibinfo{person}{Chengnian Sun}, \bibinfo{person}{Vu Le},
  {and} \bibinfo{person}{Zhendong Su}.} \bibinfo{year}{2016}\natexlab{a}.
\newblock \showarticletitle{Finding compiler bugs via live code mutation}. In
  \bibinfo{booktitle}{\emph{Proceedings of the 2016 {ACM} {SIGPLAN}
  International Conference on Object-Oriented Programming, Systems, Languages,
  and Applications, {OOPSLA} 2016, part of {SPLASH} 2016, Amsterdam, The
  Netherlands, October 30 - November 4, 2016}},
  \bibfield{editor}{\bibinfo{person}{Eelco Visser} {and}
  \bibinfo{person}{Yannis Smaragdakis}} (Eds.). \bibinfo{publisher}{{ACM}},
  \bibinfo{pages}{849--863}.
\newblock


\bibitem[\protect\citeauthoryear{Sun, Le, Zhang, and Su}{Sun
  et~al\mbox{.}}{2016b}]%
        {sunLZS16}
\bibfield{author}{\bibinfo{person}{Chengnian Sun}, \bibinfo{person}{Vu Le},
  \bibinfo{person}{Qirun Zhang}, {and} \bibinfo{person}{Zhendong Su}.}
  \bibinfo{year}{2016}\natexlab{b}.
\newblock \showarticletitle{Toward understanding compiler bugs in {GCC} and
  {LLVM}}. In \bibinfo{booktitle}{\emph{Proceedings of the 25th International
  Symposium on Software Testing and Analysis, {ISSTA} 2016, Saarbr{\"{u}}cken,
  Germany, July 18-20, 2016}}, \bibfield{editor}{\bibinfo{person}{Andreas
  Zeller} {and} \bibinfo{person}{Abhik Roychoudhury}} (Eds.).
  \bibinfo{publisher}{{ACM}}, \bibinfo{pages}{294--305}.
\newblock


\bibitem[\protect\citeauthoryear{Xiao, Liu, Yuan, Pang, and Wang}{Xiao
  et~al\mbox{.}}{2022}]%
        {DBLP:journals/pomacs/XiaoLYPW22}
\bibfield{author}{\bibinfo{person}{Dongwei Xiao}, \bibinfo{person}{Zhibo Liu},
  \bibinfo{person}{Yuanyuan Yuan}, \bibinfo{person}{Qi Pang}, {and}
  \bibinfo{person}{Shuai Wang}.} \bibinfo{year}{2022}\natexlab{}.
\newblock \showarticletitle{Metamorphic Testing of Deep Learning Compilers}.
\newblock \bibinfo{journal}{\emph{Proc. {ACM} Meas. Anal. Comput. Syst.}}
  \bibinfo{volume}{6}, \bibinfo{number}{1} (\bibinfo{year}{2022}),
  \bibinfo{pages}{15:1--15:28}.
\newblock
\urldef\tempurl%
\url{https://doi.org/10.1145/3508035}
\showDOI{\tempurl}


\bibitem[\protect\citeauthoryear{Yang, Chen, Eide, and Regehr}{Yang
  et~al\mbox{.}}{2011}]%
        {csmith}
\bibfield{author}{\bibinfo{person}{Xuejun Yang}, \bibinfo{person}{Yang Chen},
  \bibinfo{person}{Eric Eide}, {and} \bibinfo{person}{John Regehr}.}
  \bibinfo{year}{2011}\natexlab{}.
\newblock \showarticletitle{Finding and understanding bugs in C compilers}. In
  \bibinfo{booktitle}{\emph{Proceedings of the 32nd ACM SIGPLAN conference on
  Programming language design and implementation}}. \bibinfo{pages}{283--294}.
\newblock


\bibitem[\protect\citeauthoryear{Zalewski}{Zalewski}{2017}]%
        {afl}
\bibfield{author}{\bibinfo{person}{Michal Zalewski}.}
  \bibinfo{year}{2017}\natexlab{}.
\newblock \bibinfo{title}{American fuzzy lop}.
\newblock
\newblock


\bibitem[\protect\citeauthoryear{Zelenov and Zelenova}{Zelenov and
  Zelenova}{2006}]%
        {zelenov2006automated}
\bibfield{author}{\bibinfo{person}{Sergey Zelenov} {and}
  \bibinfo{person}{Sophia Zelenova}.} \bibinfo{year}{2006}\natexlab{}.
\newblock \showarticletitle{Automated generation of positive and negative tests
  for parsers}. In \bibinfo{booktitle}{\emph{Formal Approaches to Software
  Testing: 5th International Workshop, FATES 2005, Edinburgh, UK, July 11,
  2005, Revised Selected Papers 5}}. Springer, \bibinfo{pages}{187--202}.
\newblock


\bibitem[\protect\citeauthoryear{Zhao, Chen, Fu, Ye, and Wang}{Zhao
  et~al\mbox{.}}{2023}]%
        {chen2023}
\bibfield{author}{\bibinfo{person}{Yingquan Zhao}, \bibinfo{person}{Junjie
  Chen}, \bibinfo{person}{Ruifeng Fu}, \bibinfo{person}{Haojie Ye}, {and}
  \bibinfo{person}{Zan Wang}.} \bibinfo{year}{2023}\natexlab{}.
\newblock \showarticletitle{Testing the Compiler for a New-born Programming
  Language: An Industrial Case Study (Experience Paper)}.
\newblock In \bibinfo{booktitle}{\emph{The 32nd International Symposium on
  Software Testing and Analysis}}.
\newblock


\bibitem[\protect\citeauthoryear{Zhao, Wang, Chen, Liu, Wu, Zhang, and
  Zhang}{Zhao et~al\mbox{.}}{2022}]%
        {javatailor}
\bibfield{author}{\bibinfo{person}{Yingquan Zhao}, \bibinfo{person}{Zan Wang},
  \bibinfo{person}{Junjie Chen}, \bibinfo{person}{Mengdi Liu},
  \bibinfo{person}{Mingyuan Wu}, \bibinfo{person}{Yuqun Zhang}, {and}
  \bibinfo{person}{Lingming Zhang}.} \bibinfo{year}{2022}\natexlab{}.
\newblock \showarticletitle{History-driven test program synthesis for JVM
  testing}. In \bibinfo{booktitle}{\emph{Proceedings of the 44th International
  Conference on Software Engineering}}. \bibinfo{pages}{1133--1144}.
\newblock


\bibitem[\protect\citeauthoryear{Zhou, Huang, Tse, Yang, Huang, and Chen}{Zhou
  et~al\mbox{.}}{2004}]%
        {zhou2004metamorphic}
\bibfield{author}{\bibinfo{person}{Zhi~Quan Zhou}, \bibinfo{person}{DH Huang},
  \bibinfo{person}{TH Tse}, \bibinfo{person}{Zongyuan Yang},
  \bibinfo{person}{Haitao Huang}, {and} \bibinfo{person}{TY Chen}.}
  \bibinfo{year}{2004}\natexlab{}.
\newblock \showarticletitle{Metamorphic testing and its applications}. In
  \bibinfo{booktitle}{\emph{Proceedings of the 8th International Symposium on
  Future Software Technology (ISFST 2004)}}. Software Engineers Association
  Xian, China, \bibinfo{pages}{346--351}.
\newblock


\bibitem[\protect\citeauthoryear{Zhu, Hall, and May}{Zhu et~al\mbox{.}}{1997}]%
        {zhu1997software}
\bibfield{author}{\bibinfo{person}{Hong Zhu}, \bibinfo{person}{Patrick~AV
  Hall}, {and} \bibinfo{person}{John~HR May}.} \bibinfo{year}{1997}\natexlab{}.
\newblock \showarticletitle{Software unit test coverage and adequacy}.
\newblock \bibinfo{journal}{\emph{Acm computing surveys (csur)}}
  \bibinfo{volume}{29}, \bibinfo{number}{4} (\bibinfo{year}{1997}),
  \bibinfo{pages}{366--427}.
\newblock


\end{thebibliography}
\clearpage
\section*{Appendix}
\subsection*{1) Additional Examples for Cangjie}
Figure~\ref{fig:ruleCEMain} shows the implementation of the PTE rule shown in Table~\ref{tab:ruleCE}, based on the built-in AST library of Cangjie. To determine if there exists an assignment expression or a variable declaration with initialization, we enumerate each AST node of the program one by one. ``getAllNodesInTree'' is a function used to extract all the sub-nodes.
If an AST node $t$ is either `AssignExpr' (i.e., the AST node type of assignment expression) or `VarDecl' (i.e., the AST node type of variable declaration), we return `true'. Note that in the latter case, we further check whether the declared variable is initialized. Given that there could be dozens of thousands of test cases in $T$ and thousands of PTE rules, it is important that the precondition is designed such that it is fail-fast. Furthermore, some preconditions could be much more complicated than the example shown above. For instance, the implementation of the `Liskov Substitution Principle' rule requires extracting multiple nodes from the AST to implement the precondition. Concretely, we extract the name of constructors (as well as their signatures) of each class and the inheritance relation between different classes. \\
\begin{figure}[h]
    \begin{lstlisting}[basicstyle=\tiny]
        public class EqConditioanlExprStd <: PTERule{
          public override func precondition(tokens:Tokens): Bool {
          let topNode = parseFile(tokens)
          let nodes = getAllNodesInTree(topNode)
          for (node in nodes) { 
             if (node is AssignExpr) { return true } 
             if (node is VarDecl) {
                let ve = decomposeVarDecl(decl)
                if (ve.contains("initlzr")){ return true } 
             }
          }
          return false
        }
        public override func transformation(tokens:Tokens): Tokens {
              let topNode = parseFile(tokens)
              let nodes = getAllNodesInTree(topNode)
              for (i in 0..nodes.size) {
                 let node = nodes[i]
                 if (node is AssignExpr) { 
                    let leftRef = expr.getLeftValue().toTokens()
                    let rV = expr.getRightExpr().toTokens()   
                    nodes[i] =  parseAssignExpr(quote($leftRef=if(true){$(rV)}else{$(rV)}))
                 } 
                 if (node is VarDecl) { ... }   
          }
              return toTokens(nodes)
        }
          public override prop let expectations: Array<Expectation> {
              get(){ return [Expectation(Equiv)] }
          }
        }
    \end{lstlisting}
    \caption{The implementation of the rule shown in Table~\ref{tab:ruleCE}}
    \label{fig:ruleCEMain}
\end{figure}

\begin{figure}[h]
    \begin{lstlisting}[basicstyle=\tiny]
class LiskovPTE <: PTERule{
    public override func precondition(tokens:Tokens): Bool {...}
    public override prop let expectations: Array<Expectation> { ... }
    func replaceBaseName(srcExpr: CallExpr, oldBaseName: String, newBaseName: String) {
        let baseName = Token(TokenKind.IDENTIFIER, newBaseName)
        var newExpr = Tokens()
        for (tk in srcExpr.toTokens()) {
            if (tk.value == oldBaseName) { newExpr += baseName } else { newExpr += tk }
        }
        return parseCallExpr(newExpr)
    }
    public override func transformation(tokens:Tokens): Tokens{
        let inheritRelations: HashMap<String, HashSet<String>> = getInheritRelations(tokens)
        let classesInfoList: HashMap<String, ClassInfo> = getClassInfoList(tokens)
        let argTypeMap: HashMap<String, String> = getArgTypeMap(tokens)
        let topNode = parseFile(tokens)
        let nodes = getAllNodesInTree(topNode)
        for (i in 0..nodes.size) {
            let node = nodes[i]
            if (node is CallExpr) { 
               let expr = node.asCallExpr()
               let callName = expr.getBaseFunc().toTokens()[0].value
               if (inheritRelations.contains(callName)) {
                    let currentCallArgs = parseArgs(expr, argTypeMap)
                    let children = tinheritRelations[callName]
                    let candidateSubClassList = ArrayList<ClassInfo>()
                    for (childName in children) {
                        let child = this.classesInfoList[childName]
                        if (isQualifiedSubClass(child, currentCallArgs)) {
                           candidateSubClassList.append(child)
                       }
                    }
                    if (candidateSubClassList.size != 0){
                        let subClass = shuffle(candidateSubClassList)[0]
                        let newCallExpr=this.replaceBaseName(expr, callName, subClass.name)
                        nodes[i] = newCallExpr
                    }
            } 
        }
      }
     return toTokens(nodes)
   }
}
    \end{lstlisting}        
    \caption{The implementation of the Liskov Substitution Principle rule's transformation}
    \label{fig:liskovTrans}
\end{figure}

\subsection*{2) Additional Examples for Java}
Figure~\ref{fig:javabug1} shows an example of Java program that produces an unhelpful
error message.
\begin{figure}[h]
	\centering
	\begin{lstlisting}[basicstyle=\footnotesize]
Object o = null;
switch(o){
	case null: System.out.println("null");
	case Integer i: System.out.println("int: "+i); break;
	case Short s: System.out.println("short: "+s); break;
	default: System.out.println("default");
}
\end{lstlisting}
\caption{Example Java program that produces an unhelpful error message.}
\label{fig:javabug1}
\end{figure}
\end{sloppypar}
\end{document}